\documentclass{emulateapj}
\usepackage{natbib}

\def\eps@scaling{1.0}%

\newcommand\plottwotwo[2]{%
 \centering 
 \leavevmode 
 \includegraphics[width={\eps@scaling\columnwidth}]{#1}%
 \hfil 
 \includegraphics[width={\eps@scaling\columnwidth}]{#2}%
}%

\newcommand\plotthree[3]{%
 \centering 
 \leavevmode 
 \columnwidth=.67\columnwidth 
 \includegraphics[width={\eps@scaling\columnwidth}]{#1}%
 \hfil 
 \includegraphics[width={\eps@scaling\columnwidth}]{#2}%
 \hfil 
 \includegraphics[width={\eps@scaling\columnwidth}]{#3}%
}%



\long\def\comment#1{}

\def\etal{{\it et al.~}}

\def\W2{{\cal W}}

\newcommand{\tableskip}{\\[-6pt]}

\def\be{\begin{equation}}
\def\ee{\end{equation}}
\def\bea{\begin{eqnarray}}
\def\eea{\end{eqnarray}}

\def\cmm2{{\,\rm cm^{-2}}}
\def\cm2{{\,{\rm cm}^2}}
\def\cmm3{{\,{\rm cm}^{-3}}}
\def\gcmm3{{\,{\rm g\,cm^{-3}}}}

\def\fun#1#2{\lower3.6pt\vbox{\baselineskip0pt\lineskip.9pt
  \ialign{$\mathsurround=0pt#1\hfil##\hfil$\crcr#2\crcr\sim\crcr}}}

\def\ga{\mathrel{\mathpalette\fun >}}

\begin{document}

\submitted{}
\title{
Effects of sub-mm and radio point sources on the recovery
of Sunyaev-Zeldovich galaxy cluster parameters
}
\author{Lloyd Knox}
\affil{Department of Physics, 1 Shields Avenue, University of California, Davis,CA 95616}
\author{Gilbert P. Holder }
\affil{School of Natural Sciences, Institute for Advanced Study, Princeton NJ 08540}
\author{Sarah E. Church}
\affil{Stanford University Physics Department, 382 Via Pueblo Mall, Stanford, CA94305-4060}

\begin{abstract}
{Observations of clusters in the 30 to 350 GHz range can in principle
be used to determine a galaxy cluster's Comptonization parameter, $y$,
peculiar velocity, $v$ and gas temperature, $T_e$ via the dependence
of the kinetic and thermal Sunyaev-Zeldovich (SZ) effects on these
parameters.  Despite the significant contamination expected from
thermal emission by dust in high-redshift galaxies, we find that the
simultaneous determination of $\tau$, $v$ and $T_e$ is possible from
observations with sensitivity of a few $\mu$K in three or more bands with arc
minute resolution.  After allowing for realistic levels of contamination
by dusty galaxies and primary CMB anisotropy, we find that
simultaneous determinations of velocities to an
accuracy of better than 200 ${\rm km\,s^{-1}}$ and temperatures to roughly
keV accuracy should be possible in the near future.   We study how errors change
as a function of cluster properties (angular core radius and gas temperature) and
experimental parameters (observing time, angular resolution and observing frequencies).
Contaminating synchrotron emission from cluster
galaxies will probably not be a major contaminant of peculiar velocity
measurements.
}
\end{abstract}

\maketitle

\section{Introduction}

The cosmic microwave background (CMB) is a tremendous tool for studying
cosmology. The anisotropies in the CMB provide a wealth of cosmological
information (for a recent review see Hu \& Dodelson 2002)
\nocite{hu02} and upcoming experiments will provide high
sensitivity (approaching 1 $\mu K$) and high angular resolution
(approaching 1'). Such precise measurements will provide
interesting constraints on secondary anisotropies in the CMB,
imprinted by material along the line of sight at redshifts
$z \ll 1000$.

The largest secondary anisotropy is expected
to be caused by galaxy clusters, through the Sunyaev-Zeldovich (SZ)
effect \citep{sunyaev72}. Compton scattering of CMB photons
by intra-cluster electrons leaves a spatial and spectral
imprint in the CMB.  Recent measurements  of the SZ effect
have provided detailed maps of the electron distribution in
galaxy clusters (see Birkinshaw 1999 and Carlstrom \etal 2002 for
recent reviews) but new experiments offer the promise of an order
of magnitude increase in sensitivity. With $\mu K$ sensitivity
it should be possible to probe other cluster properties, such as
bulk velocities and electron temperatures.

In this paper we investigate the feasibility of measuring peculiar
velocities and cluster temperatures at cm and mm wavelengths in the
presence of contaminating sources, focusing on the effects of
dusty high-redshift galaxies. These issues have been partly addressed
previously by \citet{fischer93} and \citet{blain98} but both our
understanding of dusty distant galaxies and CMB experimental capabilities
have advanced tremendously in recent years.

We forecast errors by calculating the SZ-parameter Fisher matrix
for multi-frequency maps that contain signals from the clusters as
well as noise from CMB anisotropy, dusty galaxies and the
measurement process.  Our work extends previous studies of SZ
parameter reconstruction
\citep{haehnelt96,aghanim01,holder03,aghanim03} by
considering point source and CMB anisotropy contamination and the
influence of the atmosphere on the relative sensitivity of
different frequency bands.  While \citet{haehnelt96} did include
the effect of CMB anisotropy contamination, explicitly subtracting
it with an optimal filter, point source contamination was ignored.
And while \citet{aghanim01} included both CMB and point source
contamination, forecasts were for one specific experiment: Planck.
Here we vary experimental parameters in order to guide
experimental design.

 In \S 2 we outline the SZ effect and how it relates to
cluster properties. \S 3 summarizes the relevant known properties
of dusty galaxies observed in the direction of galaxy clusters,
while \S 4 outlines our methods for estimating the effects of
contaminating sources on cluster measurements. We discuss
complications that will arise in analyses of real data in \S 5 and
describe the experiments we consider in \S 6.  Our primary results
are in \S 7, where we forecast uncertainties for a four-channel
reference experiment as a function of cluster gas temperature and
core radius.  We explore the impact of varying experimental
parameters such as angular resolution, observing time and number
and placement of frequency channels. We follow that with a
discussion of possible radio point source contamination and close
with a discussion of our results and implications for future
instrumentation.

\section{Sunyaev-Zeldovich Effects}

The SZ effect is the change in energy of CMB photons from Compton scattering
with electrons, primarily in the intra-cluster medium. For recent reviews
see \citet{birkinshaw99} and \citet{carlstrom02}.  The high temperature
(several keV) of the electrons relative to the CMB photons leads to a net
increase of energy of the scattered photons, while
a bulk motion of the cluster electrons along the line of sight leads to
either a net redshift (moving away from observer) or blueshift
(moving toward observer) of the scattered CMB photons. The first effect
is the thermal SZ effect and it has a distinctly non-thermal spectrum;
the scattering preserves photon number and mainly just shifts the energy
of each photon, distorting the original blackbody spectrum. The effect
of the bulk motion is the kinetic SZ effect, and the emergent spectrum is
that of a blackbody with a slightly different temperature.

For the thermal SZ effect, the main physical parameters are the
optical depth to Thomson scattering, $\tau$ and the fractional
energy gain per scattering $\Theta \equiv k T_e /m_e c^2$. The
combination $\tau \Theta$ sets the amplitude of the spectral
distortion. For the kinetic SZ effect, the redshift or blueshift
is set by $\beta \equiv v/c$, so the relevant combination is $\tau
\beta$.

Typical electron temperatures are on the order of 0.01 $m_e c^2$, making
relativistic effects modestly important \citep{rephaeli95,stebbins97,
sazonov98,itoh98,nozawa98, challinor99,molnar99,dolgov01}.
The relativistic corrections
are sensitive to the electron temperature through the relativistic
correction to the Thomson cross-section and through the relativistic
form of the thermal Maxwellian velocity distribution. These corrections
can be on the order of $10\%$ or higher at many observing frequencies.

Thus the temperature difference due to the kinetic SZ effect is given by
\begin{equation}
{\Delta T_{\rm kin} \over T_{\rm cmb} } = \tau \beta
\label{eqn:kinetic}
\end{equation}
while the temperature difference due to the thermal SZ effect as a
function of frequency $\nu$ (expressed in dimensionless units
$x \equiv h\nu/k T_{cmb}$) is given by
\begin{equation}
{\Delta T_{th} \over T_{cmb} } = (x {e^x+1 \over e^x-1} -4 ) \tau
\Theta [1+\delta(x,T_e)] \quad ,
\end{equation}
where the relativistic corrections have been represented by $\delta(x,T_e)$.

Typical values for massive clusters of $\tau \sim 0.01$, $\Theta
\sim 0.01$ and $\beta \sim 0.001$ lead to typical amplitudes
($\Delta T/T_{cmb}$) for the thermal effect, kinetic effect and
relativistic corrections of $10^{-4}$, $10^{-5}$ and $10^{-5}$
respectively.

In this work, we take our fiducial galaxy cluster to have
$v = -200 {\rm km\,s^{-1}}$, $T_e = 6$ keV and $\tau = 0.01$. We vary
the thermal SZ properties using the scaling relations of
\citet{mccarthy03}. The central $y$ parameter was taken to
scale as $y \propto T^2$, which leads to $\tau \propto T$.  Self-similar
evolution would result in $y \propto T^{3/2}$; we use the slightly
steeper relation to be consistent with evidence for excess entropy.
We ignore evolution of intrinsic cluster properties with redshift, such
as the expectation that clusters (of a fixed mass) 
become more compact (and therefore hotter) at higher redshift. It is expected
that the effect of this evolution is to enhance the thermal SZ effect,
but the details of this redshift evolution are sensitive to poorly
understood gas processes \citep{holder01}.

\section{Millimeter Emission from Dusty Galaxies}

Star-forming galaxies can be very luminous in the submillimeter
wavelength range, as recent measurements by the Sub-millimeter
Common User Bolometer Array (SCUBA)
and other experiments have discovered (see Blain 2002 for a recent
review). In particular, galaxy clusters are often found to
coincide with relatively bright submillimeter emitters, probably
because of gravitational lensing effects \citep{smail02}. The
emission is mainly from dust at a temperature of several tens of
Kelvin, putting the peak of the radiation spectrum at
submillimeter wavelengths, but there can still be significant
emission at millimeter wavelengths.

Many galaxy clusters have been observed to coincide with submillimeter
sources with fluxes of several mJy at $\nu$=350 GHz. Assuming a typical
spectrum for these sources would lead to expectations of fluxes at
150 GHz on the order of 1 mJy. In a beam of 1'x1' this would correspond
to nearly 30 $\mu K$ of contamination if not correctly taken into
account. This makes dusty galaxies a non-negligible contaminant
for studies of galaxy clusters at mm wavelengths \citep{fischer93,blain98}.

Throughout we often characterize a given intensity as the
equivalent departure from the mean CMB temperature, $\Delta T$.
The conversion factor, for $\Delta T << T_{\rm CMB}$, is \bea
\left[{\partial B_\nu /\partial T}\right]^{-1} &=& \frac{c^2}{2k}\left(\frac{h}{kT_{CMB}}\right)^2\frac{\left(e^x-1\right)^2}{x^4e^x} \nonumber \\
&=& \left(\frac{119 \mu{\rm K}}{ 1 {\rm mJy}/(1' \times 1')}\right)\frac{\left(e^x-1\right)^2}{x^4e^x}
\eea

\subsection{Shot Noise}

The experiments we consider are not sensitive to the absolute
flux, but to spatial variations in the flux. These variations
contribute to the variance of temperature fluctuations on the sky,
observed with a Gaussian beam profile with full width at half maximum
of $\theta_b$,
\be (\Delta T)^2 = \sum_l {2l+1 \over 4\pi} C_l
\exp[-l^2\theta_b^2/(8 \ln{2})] \ee where \be C_l  = [{\partial B_\nu \over
\partial T}]^{-2} \int S^3 \frac{dN(>S)}{dS} d\ln S. \ee
This variance decreases with beam size as $\Delta T^2 \propto
1/\theta_b^2$.  For the convenience of using a quantity that is a
property of the sky only, and not the angular resolution of the
telescope, we define $\delta T^2 = \Delta T^2 \Omega_b$ where
$\Omega_b \propto \theta_b^2$ is the solid angle of the beam.

The shot noise at 350 GHz can be inferred from SCUBA observations.
\citet{borys03} used these observations to fit a double power-law
form for $dN(>S)/dS$: \be dN(>S)/dS =
\frac{N_0}{S_0}\left[\left(\frac{S}{S_0}\right)^a
+\left(\frac{S}{S_0}\right)^b\right]^{-1} \ee with 
$S_0 = 1.8$ mJy, $N_0 = 1.5 \times 10^4\,{\rm deg}^{-2}$, 
$a = 1.0$ and $b = 3.3.$  The result
is shot noise with an rms of $\delta T_{350} = 170 \mu$K-arcmin
where the subscript ``350'' is used to denote the frequency.

\subsection{Spectral Variations}

If the spectral behavior of these sources is known, then it would be
straightforward to measure the flux at a higher frequency, where the
emission is stronger and the SZ effects are smaller, and subtract the
appropriate levels from the measurements at other frequencies.
However, the spectral behavior is not perfectly known, so there will
be uncertainty due to imperfect subtraction, as well as uncertainty
due to noise in the measurement at higher frequencies.

To investigate the homogeneity of the spectral behavior, we used
submillimeter observations of local galaxies \citep{dunne00}.  Local
galaxies were modeled as having a spectrum of the form $I \propto
\nu^{\beta_{mm}} B_\nu$, where $B_\nu(T_{dust})$ is the blackbody
function and $\beta_{mm}$ here refers to the emissivity spectral index
(not the bulk velocity of the cluster!). Dunne \etal provide best-fit
values of $\beta_{mm}$ and dust temperatures.  We used these best-fit
spectra for the local galaxies, placed them at a range of redshifts
and fit them over the range 150 to 350 GHz assuming a power-law $I_\nu
\propto \nu^\alpha$.  As a function of redshift the mean spectral
index was well-fit by $\alpha=3.14-0.22 z$, with a fairly constant
scatter at each redshift of $\sigma_\alpha \sim 0.18$.  The moderate
flattening with increasing redshift arises because the radiation that
is observed at 350 GHz originates at a higher frequency, closer to the
peak of the dust emission.

Without source redshifts, it is not clear what to take as the
mean spectral index.  Assuming a uniform distribution
in redshift between 0 and 5, the mean spectral index  is 2.6 and
the $rms$ scatter around this mean is 0.4. The sources are unlikely
to be at very low redshift, simply due to volume considerations,
but the redshift distribution of the observed submillimeter-emitting
galaxies is very uncertain.  We set our mean value for $\alpha$ as
$\alpha = 2.6$ and set $\sigma_\alpha = 0.3$, unless specified otherwise.
We set $\sigma_\alpha < 0.4$ since the redshift distribution will be
more peaked than the uniform one that leads to 0.4.

The quantity $\sigma_{\alpha}$ is the spectral index rms for {\em
individual} galaxies.  For $N$ galaxies of similar brightness the
composite intensity will have a scatter in spectral index that is
smaller by $\sqrt{N}$.  We define an effective number density of
galaxies by weighting them according to their contribution to
shot-noise variance.  Thus, \bea N_{\rm eff} &=& \int S^3
\frac{dN^2(>S)}{dS} d\ln{S}/\int S^3 \frac{dN(>S)}{dS} d\ln{S}
\nonumber \\ &=& 0.64/({\rm sq. arcmin}) \eea where the last
equality follows from the \citet{borys03} $dN(>S)/dS$.  For
simplicity we set $N_{\rm eff} = 1/$(sq. arcmin).  Finally we
define $\bar \sigma_\alpha = \sigma_\alpha/\sqrt{N_{\rm eff}}$ and
take our fiducial value to be $\bar \sigma_\alpha = 0.3$-arcmin.

\subsection{Gravitational Lensing}

The \cite{borys03} $dN(>S)/dS$ is for an unlensed population of
sources.  
Lensing changes both
the brightnesses of galaxies and their number densities.  Indeed, the
magnification provided by galaxy clusters has been exploited to study
the population to fainter limiting magnitudes \citep{smail97,smail02}.  
The magnitude, and even the sign, of the effect on shot noise depends on
$dN(>S)/dS$.  Magnification increases $S$ which increases shot noise,
but also decreases $dN(>S)/dS$ (at fixed unlensed $S$) which decreases
shot noise.  For a power-law, $dN(>S)/dS \propto S^\alpha$, the net effect
is an increase in shot noise if $\alpha < -2$.
For the relatively bright population at 350 GHz of interest for this work, 
the shape of $dN/dS$ is such that
lensing leads to an enhancement of shot noise.

\citet{blain98} pointed out that lensing would significantly enhance
the level of contamination by IR luminous galaxies of SZ observations.
The key quantity for lensing is the Einstein radius, $\theta_E$.  A
point source with source plane location coincident with the cluster center
($\theta = 0$) will form a ring of infinite magnification appearing in
the image plane at distance $\theta_E$.  In the image plane,
magnification rises from the center towards $\theta_E$ and then drops
beyond $\theta_E$.  Thus, as \citep{blain98} showed, the enhancement
of confusion noise for an observation towards the cluster center
increases with increasing beam size until peaking at $\theta_b =
2\theta_E$ as more weight is placed on galaxies near $\theta_E$, and
then drops with further increase in $\theta_b$ as the effect is
diluted by adding in more galaxies from beyond the Einstein radius.

For a rough model of the effect of lensing we introduce the
confusion noise enhancement factor, $E(\theta_b,\theta)$ which is
the ratio of lensed confusion noise to unlensed confusion noise at
distance $\theta$ from the cluster center and for observations
with angular resolution $\theta_b$.  For the dependence of $E$ on
$\theta_b$ we use the results of \citet{blain98} for his galaxy
evolution model I1 for a cluster with velocity dispersion of
$\sigma_v = 1360$ km/s and $\theta_E= 16''$ when it is placed at
$z=0.171$.  For this case $E(\theta_b,0) =$ 1.1, 1.5, 2.5, 2.1,
1.8 and 1.5 for $\theta_b = $ 10, 20, 40, 60, 100 and 200 arcsec
respectively.  We ignore the redshift-dependence of $E$ since for
$\theta_b \ga 30''$ \citet{blain98} shows that $E$ is highly
independent of the cluster redshift.  We also ignore the
$\sigma_v$ dependence of $E$ since \citet{blain98} shows this
dependence to be quite slow, covering a range of 30\% as
$\sigma_v$ varies from 800 km/s to 2000 km/s. For $\sigma_v =
1360$ km/sec one expects a gas temperature of about $T = 10$ keV
\citep{girardi96}. For dependence on $\theta$ we simply set
$E(\theta_b,\theta) = E(\theta_b,0)$ for $\theta < \theta_b$ and
$E(\theta_b,\theta) = 1$ otherwise. For all cases that we consider
here the beam size will be larger than a typical Einstein radius
for a cluster. 

\begin{figure}
\plotone{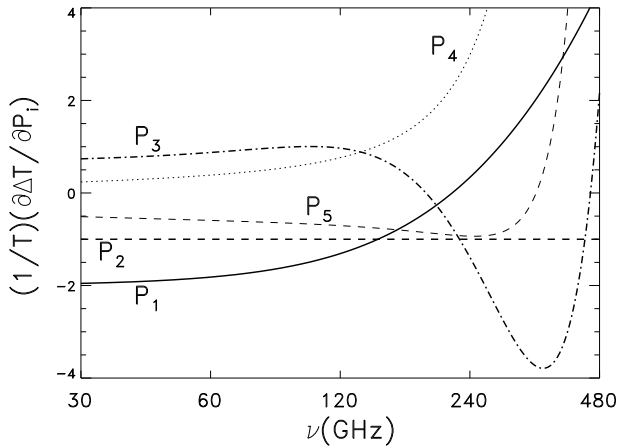}
\caption{ Derivatives of the frequency--dependent temperature
fluctuation data, $\Delta T$, with respect to the parameters $P_1 = y
= \tau \theta$, $P_2 = \tau \bar \beta$, $P_3
= \tau \theta^2$, $P_4 =\Delta T_{350}/T_{CMB}$ and $P_5 = \alpha$
(see text).  The $P_3$ derivative has been divided by 5 to fit on the
plot and the $P_5$ derivative is arbitrarily normalized.  }
\label{fig:derivatives}
\end{figure}

\section{Methods}

We model the data at sky location $\theta_i$ and frequency $\nu_a$ as having
contributions from both signal and noise:
\be
\frac{\Delta T}{T_{\rm CMB}}\left(\nu_a,\theta_i \right)  =  s(\nu_a,\theta_i) +
n(\nu_a,\theta_i).
\ee
We consider each component in detail below.

\subsection{Signal Model}

The signal is due to three components,
\be
s(\nu_a,x_i) = \sum_{m=1}^3 P_m \tilde h(x_i)f_m(\nu_a),
\label{eqn:signals}
\ee
where
\bea
P_1 &\equiv & y = \int dl \sigma_T n_e \Theta = \tau \bar \Theta_\tau\\
P_2 &=& \int dl \sigma_T n_e v/c  =
\tau \bar \beta_\tau \ \ {\rm and} \\
P_3 & =& \int dl\sigma_T n_e \Theta^2 = \tau \bar {\Theta_\tau^2} = y \bar 
\Theta_y.
\eea
Here $n_e$ is the electron number density and $\Theta \equiv kT_e/(m_e c^2)$
is the dimensionless gas temperature.  The subscripts on the average
quantities indicate the weighting along the line of sight.  For $\tau$ the
weighting is with the electron-number density and for $y$ the weighting
is with the pressure, $n_e \theta$.  The optical depth, $\tau$ is given by
$\tau = \int dl \sigma_T n_e$.

The integrals defining $P_1,P_2$ and $P_3$ are along the line of sight
through the cluster center.
We take the angular profile of the cluster to be
\be
h(\theta) = \left(1+\left(|\theta|/\theta_c\right)^2\right)^{-1/2}
\ee
The beam-convolved cluster profile is denoted by $\tilde h(\theta)$,
and we assume Gaussian beams in all that follows. Note that the
assumption of the same cluster profile for all three signal
components in equation \ref{eqn:signals} is only valid for an
isothermal intracluster medium.

The frequency dependences are given by \citep{itoh98}
\bea
f_1(\nu) & \equiv & \tilde x -4 \\
f_2(\nu) & \equiv & 1 \\
f_3(\nu) & \equiv & -10+\frac{47}{2} \tilde x
- \frac{42}{5} \tilde x^2 + \frac{7}{10} \tilde x^3 +
\Bigl(-\frac{21}{5}+\frac{7}{5} \tilde x\Bigr)\tilde s^2
\eea
where
\bea
\tilde x &\equiv & x {\rm coth}(x/2) \\
\tilde s &\equiv & x/{\rm sinh}(x/2).
\eea
The first and third components are due to the scattering of CMB
photons off of the hot electrons in the cluster and represent
the thermal SZ effect, while the second component is the kinetic
SZ effect.
We have neglected
contributions to the SZ effects that are $\tau$ times higher order
products of $\beta$ ($\sim 10^{-3}$) and $\theta$ ($\sim 10^{-2}$).
For clusters with temperatures of 10 keV or lower the higher order
corrections are less than five percent
\citep{challinor98,stebbins97,itoh98} at all frequencies except near
the null of the thermal SZ spectrum (where the thermal SZ signal
is sub-dominant).

Because they are the amplitudes of the different frequency shapes,
$P_1$, $P_2$ and $P_3$ are the theoretical parameters most
directly related to the data.  From them one can estimate the
physically interesting quantities:
\bea 
\hat \Theta_y & = & P_3/P_1 = \bar \Theta_y \\ 
\hat \tau & = & P_1^2/P_3  =  
\tau \frac{\left(\bar \Theta_\tau \right)^2}{\bar {\Theta_\tau^2}}  =  
\tau \frac{\bar \Theta_\tau}{\bar \Theta_y} \\ 
\hat \beta & = & P_2/{\hat \tau}   = {\bar \beta}
\frac{\bar \Theta_y}{\bar \Theta_\tau} \eea 
Thus the temperature determinations from SZ measurements are measurements
of the pressure-weighted average gas temperature.  Velocity measurements
are of the mass-weighted average velocity (assuming electron
density traces mass) times a correction factor dependent on differences
in different temperature averages.  The optical depth determinations
are of the optical depth times another ratio of different 
average temperatures.  Throughout, we assume isothermality
in order to avoid these complications; the distinctions must be
kept in mind when analyzing real data, particularly for comparison of
X-ray and SZ gas temperatures.

In a recent preprint \citep{hansen04} it was noted that peculiar velocity
measurements of non-isothermal clusters could be biased. The source of this
bias can be clearly seen in the above expression for $\hat \beta$, where
the true $\bar \beta$ is multiplied by the ratio of two different weightings
of the electron temperature. In general the optical depth weighted
temperature will not be the same as the pressure weighted temperature,
resulting in a biased velocity estimate. This is a direct consequence
of the biased estimate of the optical depth.

\subsection{Noise Model}

The noise has a contribution from measurement error, $n_{\rm
inst}$, from primary CMB fluctuations, $n_{\rm CMB}$ and from
emission from galaxies in the direction of the cluster $n_{\rm
gal}$. We assume the composite dusty galaxy spectrum has a
power-law form in intensity, $I_\nu \propto \nu^\alpha$ where
$\alpha$ is spatially varying with $\alpha(\theta) = \bar \alpha
+\delta \alpha(\theta)$ and $\bar \alpha= 2.6$ as discussed
earlier. Linearizing the dependence of $I_\nu$ on $\delta \alpha$
we can write $n_{\rm gal}$ as \be n_{\rm gal}(\nu_a,\theta_i) =
\frac{\delta T^{\rm gal}_{350}}{T_{\rm CMB}}\left(\nu_a,\theta_i
\right)\left[f_4(\nu_a)+\delta \alpha(\theta_i)
f_5(\nu_a)\right]E(\theta_i,\theta_b) \ee where
$E(\theta_i,\theta_b)$ is the shot noise enhancement factor
discussed earlier.  The frequency dependences are given by \bea
f_4(\nu) &=& \nu^{\bar \alpha-2}
x^{-2} e^{-x}\left(e^x-1\right)^2/f_4(350 \ {\rm GHz}) \\
f_5(\nu) &=& \ln{\left({\nu \over 350\ \rm{ GHz}}\right)} f_4(\nu)
\eea

The CMB noise contribution is independent of frequency, and the
instrument noise contributions as a function of frequency depend
on the experiment being considered.  The total noise in the map is
then $n_{\rm tot} = n_{\rm CMB}+n_{\rm gal}+n_{\rm inst}$. The
noise in the map therefore has the two-point function, considering
points in the map $\theta_i$ at frequency $\nu_a$ and $\theta_j$
at frequency $\nu_b$:
\begin{eqnarray}
C^{noise}_{i a j b}  & \equiv & \langle n_{\rm tot}
  (\theta_i,\nu_a)n_{\rm tot}(\theta_j,\nu_b)\rangle
 \\ \nonumber
 & = & C^{\rm CMB}_{i j}+
\\ \nonumber   & & 
\hskip -0.5in  
\langle \left(\delta T_{350}^{\rm gal}/T_{CMB}\right)^2\rangle /\Omega_{\rm b} E(\theta_i,\theta_b)E(\theta_j,\theta_b)
c_{i j }^{PS} f_4(\nu_a)f_4(\nu_b)  
\\ \nonumber & &
\hskip -0.5in  
+ \sigma_\alpha^2 \langle \left(\Delta T_{350}/T_{CMB}\right)^2\rangle /\Omega_{\rm b}^2 E(\theta_i,\theta_b)E(\theta_j,\theta_b)
\left[c_{i j }^{PS}\right]^2 f_5(\nu_a)f_5(\nu_b)
\\ \nonumber
 & &+\sigma^2(\nu_a)/\Omega_{\rm pix}\delta_{ab}\delta_{i j}
\end{eqnarray}

In the above, $c^{PS}$ is a shot-noise covariance matrix (with
off-diagonal correlations only due to beam-smoothing) normalized
to unit variance, and the last term gives the contribution of
instrument noise, which will be considered in detail in \S 6. 
$C^{\rm CMB}$ is the covariance matrix of the
CMB fluctuations, given by \be C^{\rm CMB}_{i j } = \sum_l
\frac{2l+1}{4\pi} C_l P_l(\cos \theta_{i j })
\exp\left[-l^2\theta_b^2/(2\ln 2)\right] \ee where $\theta_{i j}$
is the angular separation between pixels $i$ and $j$.

This analysis implicitly assumes that we know the statistical
properties of all sources of noise, including the dusty galaxies.
This is an excellent approximation for the CMB fluctuations, but the
dusty galaxy noise power spectrum will be affected by both
clustering and gravitational lensing. On arcminute scales the clustering
effects should be relatively small, but the effects of gravitational
lensing will modify the noise properties.

\subsection{Error Forecasting}

We forecast how well the three parameters of our model can be
measured by calculating a Fisher matrix.  Combining the spatial
and spectral indicators $i$ and $a$ into a combined index $\mu$
(and $j$ and $b$ into $\nu$) we can write the Fisher matrix as \be
F_{pp'} = \frac{1}{T_{\rm CMB}^2} \frac{\partial \Delta
T_\mu}{\partial P_p}\left(C^{noise}\right)^{-1}_{\mu \nu}
\frac{\partial \Delta T_\nu}{\partial P_{p'}} \ee To this Fisher
matrix we usually add a prior Fisher matrix which carries the
information we have that is not contained in our measurements in
the 30 to 350 GHz range. The error covariance matrix for our three
parameters is then given by \be C_{pp'} = \left[\left(F+F_{\rm \rm
prior}\right)^{-1}\right]_{pp'}. \ee Note that since our data
depend linearly on our parameters $P$, this Fisher matrix
calculation of the expected error covariance matrix is exact,
given our model of the data.

We include prior information due to X-ray measurements of the
gas temperature with error
$\sigma_{T_x} = \sigma_\theta \times 511\ {\rm keV}$ so that
\bea
F_{\rm prior}(1,1) &=& \left(\sigma_\theta \tau\right)^{-2} \nonumber \\
F_{\rm prior}(3,3) &=& \left(\sigma_ \theta y\right)^{-2} \ {\rm and} \\
F_{\rm prior}(1,3) = F_{\rm prior}(3,1) &=& -\sqrt{F_{\rm
prior}(1,1)F_{\rm prior}(3,3)}. \nonumber \eea Including the prior
information in this manner is approximate since it results from
Taylor expanding the dependence of $\theta$ on $P_1$ and $P_3$.

To calculate errors on a new set of parameters, $\tilde P$, that
are functions of our original parameters $P$, we Taylor expand the
dependence of $\tilde P$ on $P$ to first order about the fiducial
value.  We then calculate the Fisher matrix in these new
coordinates by: \be \tilde F_{ii'} = \sum_{pp'}
R_{ip}F_{pp'}R^T_{p'i'} \ee where the transformation matrix
$R_{ip} = \partial P_p/\partial \tilde P_i$.  For example, such a
variable transformation is necessary to get errors on $\Theta =
P_3/P_1$.  Dropping the higher-order terms in the Taylor expansion
makes this procedure approximate as well.

Calculating these Fisher matrices can be
computationally expensive as
matrices of size $N_{\rm pix} N_{\nu}$ have to be inverted and
multiplied where $N_{\rm pix}$ is the number of map pixels
and $N_\nu$
is the number of frequencies.  We therefore wish to pixelize no more
finely than necessary, and keep the physical area of the map as small
as possible.  If the map is too small compared to the beam-convolved
cluster profile we will lose the large-scale information necessary for
subtracting off the CMB contamination.  If the pixel size is too large
we will lose small-scale information.  We have found the following
prescriptions to ensure we are neither losing information, nor
using much larger $N_{\rm pix}$ than necessary:
\bea
\theta_M &=& 4(2\theta_c + \theta_b) \nonumber \\
\theta_{\rm pix} &=& (\theta_c+\theta_b)/2.5
\eea
where the map and pixels are squares of length
$\theta_M$ and $\theta_{\rm pix}$ respectively.

\section{Possible Real-World Complications}

Our model is fairly detailed and sufficient in many ways, but there
are several respects in which real clusters could present some
challenges, such as temperature structure in the intra-cluster medium,
incomplete knowledge of the lensing properties of galaxy clusters,
and internal bulk flows in the cluster.

Temperature gradients in the intra-cluster medium would lead to a
mis-match between the signal maps of equation \ref{eqn:signals}.
Observed clusters show significant departures from isothermality
\citep{degrandi02} at relatively large radii and at small radii,
but the largest scales are already obscured by the primary CMB
anisotropies and the small scales contribute relatively little to
the SZ effect. It is therefore not expected to be a significant
effect but will complicate data analysis.

Similarly, we have assumed that we know the cluster template.
In the cases where we assume complementary X-ray spectroscopy it
is quite reasonable to assume that the X-ray image gives an
estimate of the electron spatial distribution. Without X-ray
information the best estimate of the spatial template will come
from the thermal SZ map itself, where a large mismatch between
the assumed spatial template and the observed emission should be
evident. A parametrized model for the cluster could
be introduced into the fit without significantly affecting the
constraints on peculiar velocities.

We have assumed that the magnification due to lensing is a simple
step function in radius. The main effect of lensing is to amplify
the Poisson noise, and the amount of amplification depends on the
details of the mass profile. Incomplete knowledge of the mass distribution
will therefore lead to less efficient component separation.
Deep optical images will be useful for the purpose of studying
the strong lensing properties of galaxy clusters.

On a related note, we have assumed that we know the statistical
properties of the galaxy contamination. Lensing modifies the noise
properties of the background galaxies, and there is uncertainty in
the statistics of dusty galaxies. Multi-frequency observations of
many fields, both with and without clusters, will lead to a good
understanding of the statistics of dusty galaxies.

We have also assumed a single peculiar velocity for the cluster whereas
real clusters show evidence for internal flows, sometimes
as large as 3-4000 km s$^{-1}$ \citep{dupke02, markevitch03}.
Some striking visualizations of the kinetic SZ fluctuations induced by
these internal flows are given in \citet{nagai03}.
It has been shown \citep{haehnelt96,holder03,nagai03}
that the average peculiar velocity provides an unbiased estimate of the
true bulk velocity but with an added dispersion of roughly
100 km\,s$^{-1}$.  For our purposes this can be considered as an extra
source of noise which is small compared
to the uncertainties due to astrophysical confusion.

We have neglected calibration uncertainty.  Since the total signal,
anywhere other than near the thermal SZ null, is about ten times as
large as the kinetic SZ signal, accurate calibration of one band
relative to another is important.  Calibration errors will have to be
controlled to better than about 10\% for peculiar velocity
measurements to be better than about one $\sigma$.  The same argument
applies to gas temperatures for clusters with temperatures near 6 keV,
since the relativistic correction is also about 10\% of the total signal.
But gas temperatures, since they do not suffer from CMB anisotropy
confusion, can be measured to much better than 1 $\sigma$.  Achieving
$\pm 1$ keV, as can be done if we neglect calibration uncertainty, may
require calibration uncertainties as small as 2\%.  Measurements
closer to the thermal SZ null will reduce these sensitivities to
calibration error.

\section{Experiments}

As our reference experiment we have modeled the sensitivity
achievable for a generic ground-based bolometric experiment with
three frequency bands selected to lie in the atmospheric windows
available at a good millimeter site. We have also assumed that a
30 GHz measurement is available, with sensitivity similar to that
expected from the SZA. We assume that the sensitivity in each
channel of the bolometric experiment is limited only by
fluctuations in the photon background itself and that other
potential sources of noise, such as phonon fluctuations in the
bolometers or electronic readout noise, are negligible (this can
be achieved with current detector technology, see Lange 2002).
\nocite{lange02}
We assume that $1/f$ noise introduced by fluctuations in
atmospheric water vapor emission can be adequately subtracted from
all of the data. This can be accomplished by some kind of spatial
chopping or removal of common-mode signals from array data, or
alternatively the spectral properties of the atmospheric noise can
be used.  The latter technique has been demonstrated by the
Sunyaev-Zeldovich Infrared Experiment (SuZIE) which operates at
150, 220 and 350 GHz and which has been used to set limits to the
peculiar velocities of 11 galaxy clusters \citep{benson03a,
benson03b}. For the purposes of this paper we do not include in our calculations any of
the overhead (increase in real observation time) that this process will
produce, since this will depend on the individual experiments.  For
example, since the atmosphere is in the near field of most large
telescopes, many detectors view the same column of atmosphere, yet observe
different parts of the cluster, or none of the  cluster at all.  By using
detectors that are a large distance from the cluster to subtract
common-mode fluctuations, it may be possible to subtract atmospheric noise
with little or no degradation in sensitivity.  This is a commonly-made
assumption that we also include here.

With the above assumptions the performance at each frequency is
characterized by the noise equivalent power (NEP; quoted in
W\,Hz$^{-1/2}$), given by:
\begin{equation}
  {\rm NEP}^2=2P_{\rm load}h\nu+ \frac{P_{\rm
  load}^2}{n\Delta\nu}
  \label{model:e1}
\end{equation}
where $P_{\rm load}$ is the total background loading on the
detector in Watts, $\Delta\nu$ is the detection bandwidth in GHz
and $n$ is the number of waveguide modes that are detected ($n=1$
for a diffraction-limited system). The first term in
equation~\ref{model:e1} is caused by shot noise due to Poisson
statistics of the incoming photons while the second term accounts
for the effect of photon (boson) correlation. For more detail, see
\cite{lamarre86}. Sources of background loading include the
atmosphere, warm emission from the telescope and surroundings, and
of course the CMB itself.  The total background loading is then:
\begin{equation}
P_{\rm load} = \sum\limits_{i} P_i
\label{model:e2}
\end{equation}
where the sum is over all sources of power on the detector. For
each source of loading, the power at the detector can be
calculated from:
\begin{equation}
  P_i  = \int \frac{2kT_{{\rm RJ},i}\,\nu^2}{c^2}\, \eta_i \, A\Omega_b {\rm d}\nu
  \label{model:e3}
\end{equation}
where  $T_{\rm RJ, i}$ is the equivalent Rayleigh-Jeans (RJ)
temperature of the $i$th background source, $\eta_i$ is the
fraction of photons from the background source that reach the
detector, $A$ is the effective telescope area and $\Omega_{\rm b}$
is the solid angle response of each detector on the sky. The
integral is over the spectral band to which the detector responds.
We assume for simplicity, a square band of width $\Delta\nu$. We
now consider each source of background power in turn.

For the telescope the equivalent RJ temperature $T_{\rm RJ, 1}$
is:
\begin{equation}
  T_{\rm RJ, 1} = \epsilon(\nu) T_{\rm tel} \frac{x_L}{e^{x_L}-1}
  \label{model:e4}
\end{equation}
where $T_{\rm tel}$ is the physical temperature of the telescope,
$\epsilon$ is a frequency-dependent emissivity and
$x_L=h\nu/kT_{\rm tel}$.

For the atmosphere:
\begin{equation}
  T_{\rm RJ, 2} = T_{\rm atm}\left\{1-\exp[-\tau(\nu)/\cos\theta_{\rm za}]\right\}
  \label{model:e5}
\end{equation}
where $T_{\rm atm}$ is the physical temperature of the telescope,
$\tau(\nu)$ is the frequency-dependent zenith optical depth and
$\theta_{\rm za}$ is the zenith angle of the observation.

For the CMB:
\begin{equation}
  T_{\rm RJ, 3} = T_{\rm CMB} \frac{x}{e^x-1}
  \label{model:e6}
\end{equation}
where $T_{\rm CMB}=2.726$\,K and $x=h\nu/kT_{\rm CMB}$.

The optical efficiency is assumed to be:
\begin{equation}
\eta_i = \left\{ \begin{array}{ll}
  0.4 & \mbox{for $i=1,2$ (Tel, Atm) } \\
  0.4 \times \exp[-\tau(\nu)/\cos\theta_{\rm za}]  & \mbox{for $i=3$ (CMB)}
\end{array} \right.
\label{model:e7}
\end{equation}
which are values typical of those measured for
bolometric systems \citep{holzapfel97, runyan03}.

Once the NEP is determined, the sensitivity to
CMB fluctuations,
including the SZ effect, is obtained by calculating the noise
equivalent temperature (NET) as follows:
\begin{equation} {\rm NET} = \frac{1}{\sqrt{2}\eta_{\rm CMB}}
  \frac{{\rm NEP}}{(\partial P_{\rm CMB}/\partial T_{\rm CMB})} \mbox{\hspace*{0.5cm} K s$^{-1}$}  \quad ,
  \label{model:e8}
\end{equation}
where $\eta_{\rm CMB}$ is given above.

We further assume that the focal plane is fully sampled with detectors
spaced at half the beam size.  Each detector will have an angular response
that is
diffraction-limited at each frequency, in which case the solid
angle response of each detector is just $\Omega_{\rm
b}=c^2/(A\nu^2)$. With a suitable scanning strategy, an oversampled map can
always be produced with a pixel solid
angle $\Omega_{\rm p}$.  The rms sensitivity per map pixel is then:
\begin{equation}
\delta T/\sqrt{\Omega_{\rm p}} = \Delta T = \frac{{\rm NET}}{f}
\times \sqrt{\frac{\Omega_{\rm b}}{\Omega_{\rm p} t_{\rm int}}}
\label{model:e9}
\end{equation}
where $t_{\rm int}$ is the integration time per detector and the NET is per 
focal plane element. The factor $f$ accounts for the increase in 
sensitivity that a fully sampled focal plane can achieve over a focal plane 
with feed horns that are exactly diffraction-limited. We assume $f=1.58$ 
which is the ideal limit.  The real value of $f$ depends upon factors such 
as detector noise and instrument photon background \citep{griffin02}. 

We include a map pixel size here to allow us to compare 
experiments with different angular resolutions. If the array is large 
enough the scanning can be arranged so that there are always focal plane 
elements viewing the cluster. No integration time is lost in this case. 
Other chopping schemes could degrade the sensitivity by a factor that 
depends on the details of the particular scheme being used.  Thus our final 
numbers may be taken as best case values.

\begin{figure}
\begin{center}
\resizebox{3.5in}{!}{\includegraphics{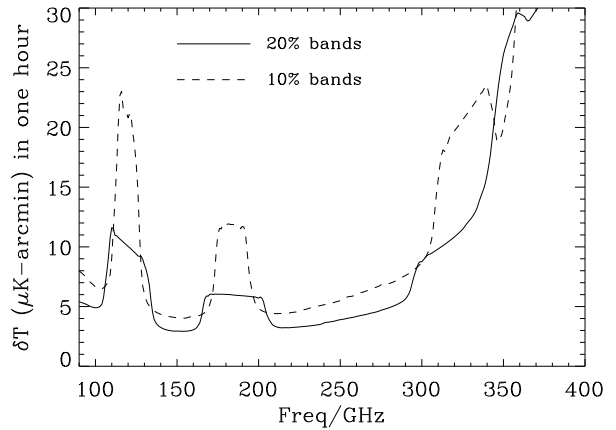}}
\end{center}
\caption{\footnotesize%
Error in each arcminute pixel for an hour of
integration with our reference experiment as a function
of central frequency, for bandwidths of 20\% (solid line)
and 10\% (dashed line).}
\label{model:f1}
\end{figure}

In order to use equations~\ref{model:e1} through \ref{model:e9}, 
we now define the details of the experiments.  For our reference 
experiment we assume a ground-based off-axis 8-m telescope located 
at a site with an altitude of 17,000$'$ and precipitable water 
vapor of 0.5\,mm. The telescope and instrument
are assumed to have an equivalent Rayleigh-Jeans temperature of
$T_{\rm RJ, 1}=10$\,K that is frequency independent.
Figure~\ref{model:f1} shows the dependence of $\delta T$ on
frequency, for a map pixel size of 1 sq arcmin. Two models are
shown, one with a bandwidth of 20\% at each frequency and one with
a bandwidth of 10\%.  There are broad atmospheric windows visible
at 135-165\,GHz and at 215-290\,GHz. The window at 345\,GHz is
sufficiently narrow that only a 10\% wide frequency band can be
utilized but it is a very useful atmosphere and/or point source
monitor. Based on this figure we have placed three of the four
channels of the reference experiment at 150, 220 and 280\,GHz. The
sensitivities are shown in Table~\ref{tab:experiments}.

\begin{table}
\begin{center}
\begin{tabular}{cccc}
Experiment & $\nu$  & $\theta_b$ $^1$ & $\delta T$ $^2$\\
           &  (GHz) &    (')     &  ($\mu$K-arcmin)    \\
\tableskip\hline \tableskip
Reference   &  30  & 1.0 &  7.0 \\
 (1 hr)     &  150 & 1.3 &  2.8 \\
            &  220 & 0.9 &  3.1 \\
            &  280 & 0.7 &  4.6 \\
\tableskip\hline
SZA$^3$         &   30 & 1.0  & 10.0 \\
\tableskip\hline
SuZIE-III$^4$   &  150 & 1.0 &  9.2 \\
            &  220 & 0.7 &  15 \\
            &  280 & 0.5 & 28 \\
            &  345$^5$ & 0.5 & 110 \\
\tableskip\hline
ACT$^6$     &  145 & 1.7 &  3.4 (3.4)\\
            &  225 & 1.1 &  3.7 (3.4)\\
            &  265 & 0.9 &  4.3 (3.4)\\
\tableskip\hline

\end{tabular}
\end{center}
$^1$ Resolutions assumed for our forecasting, for simplicity, are 1'
for all experiments and all channels, except for ACT where we assume
1.7' for all channels.\\
$^2$ The sensitivity numbers are per 1' pixel assuming
$f=1.58$.\\
$^3$ Sensitivity for mapping an individual cluster in 12 hours (Leitch
private communication)\\
$^4$ Sensitivity achievable in one hour over a 48 sq. arcmin map.\\
$^5$ This channel is unused in our forecasting.  We assume it is used
entirely for atmospheric subtraction.\\
$^6$Numbers in parentheses are from http://www.hep.upenn.edu/\~angelica/act/act.html for
a 100 sq. degree map.  These are the numbers we use in our forecasting.
\caption{Assumed experimental specifications and sensitivities
per square arcminute pixel. }

\label{tab:experiments}
\end{table}

The Table also shows sensitivities for several experiments
currently being built. The Sunyaev-Zeldovich Array (SZA,
\citet{carlstrom02}) is a 30 GHz interferometer which will begin
operation in early 2004. The SZA will also be able to observe
clusters with higher angular resolution, but lower sensitivity, at
90 GHz (not shown in the Table). SuZIE III will operate at the
Caltech Submillimeter Observatory (CSO) and will observe a $6'
\times 2'$ field of view simultaneously at 150, 220, 280 and 350
GHz with four 48-pixel arrays. SuZIE III has somewhat worse
sensitivity than the reference experiment because the CSO is an
on-axis Cassegrain telescope, leading to higher telescope loading
due to the secondary mirror support legs obstructing the beam. The
Atacama Cosmology Telescope (ACT) is a proposed 6m telescope to be
sited on the Atacama Plateau in Chile. The ACT will map one
hundred square degrees of sky simultaneously in three frequency
bands with $32\times32$ element arrays and is expected to detect
large numbers of clusters through the SZ effect.

\section{Results}

Here we display and discuss our error forecasts.  Many of our
plots show error forecasts as a function of the galaxy cluster
parameters $\theta_c$ and $T_e$.  We use the core radius
as a proxy for redshift; for a fixed three-dimensional 
gas distribution the central $y$ value is redshift-independent,
so the redshift-dependence of the signal comes entirely from 
$\theta_c \propto 1/D_A(z)$ where $D_A(z)$ is the angular diameter
distance to the cluster.  The core radius is also an interesting
parameter since it has a large influence on the peculiar velocity
error \citep{haehnelt96}.  Cluster mass is related to gas temperature by 
$M = 8\times 10^{14} h^{-1}M_{\odot} (T/6 {\rm keV})^{3/2}$ 
(although the normalization is not well-established)
so the temperature range 
3 to 15 keV corresponds to a mass
range 
$3\times 10^{14}$ to $ 3 \times 10^{15}\ h^{-1} M_{\odot}$.

We study first the dependence of our error forecasts 
on galaxy cluster parameters and point source parameters.  
Then we look at the dependence on experimental parameters.
Finally, we forecast errors for some planned experiments.  We
vary many parameters and each time hold many other parameters fixed.
Table ~\ref{tab:fiducial} shows the values of parameters we use
unless otherwise specified.  For the reader's convenience, this is a
redundant list; for example $\Theta$ is specified as well as $T$
even though $\Theta$ is simply $kT/(m_e c^2)$.

\begin{table*}
\begin{center}
\begin{tabular}{|c|c|c|c|c|c|c|c|c|c|c|c|c|c|c|c}
\hline 
$z$ & $\tau$ & $T$  & $\bar \Theta$ & $y$ & $v=\bar \beta c$ & $\bar \theta_c$ 
& $\bar \alpha$ & $\bar \sigma_\alpha$ & $\langle (\delta T^{\rm gal}_{350})^2\rangle^{1/2}$ & $\theta_b$ & $t_{\rm obs}$ \\
\hline 
0.5   &  0.01  & 6 keV &  $0.012 $ & $1.2\times 10^{-4}$ & -200 km/s & 30'' 
& 2.6 & 0.3-arcm & 170$\mu$K-arcm & 60'' & 1 hr\\
\hline
\end{tabular}
\end{center}
\caption{Fiducial parameters.}
\label{tab:fiducial}
\end{table*}

\begin{figure}
\plotone{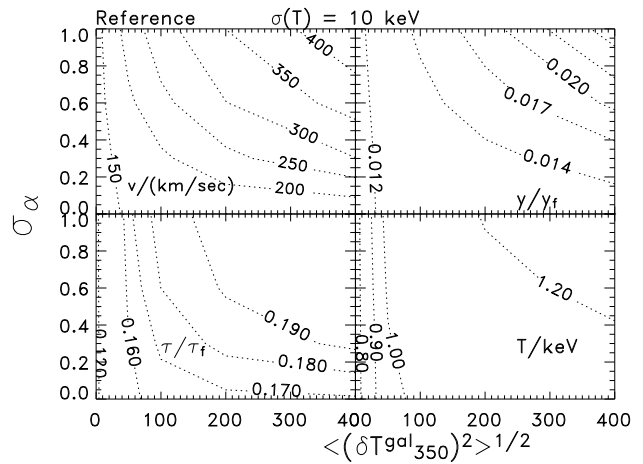} \caption{
Forecasted uncertainties for one hour of observation with the Reference 1', 4-band experiment
and fiducial cluster parameter values of $\theta_c = 30''$ and $T
= 6$keV as a function of confusion noise rms in units of
arcminutes and spectral index rms in units of $\mu$K-arcmin.
Fiducial point source parameter values are $\bar \sigma_\alpha =
0.3'$ and $\langle \left(\delta T^{\rm
gal}_{350}\right)^2\rangle^{1/2} = 170 \mu$K-arcmin. }
\label{fig:dusty}
\end{figure}

\begin{figure*}
\plottwotwo{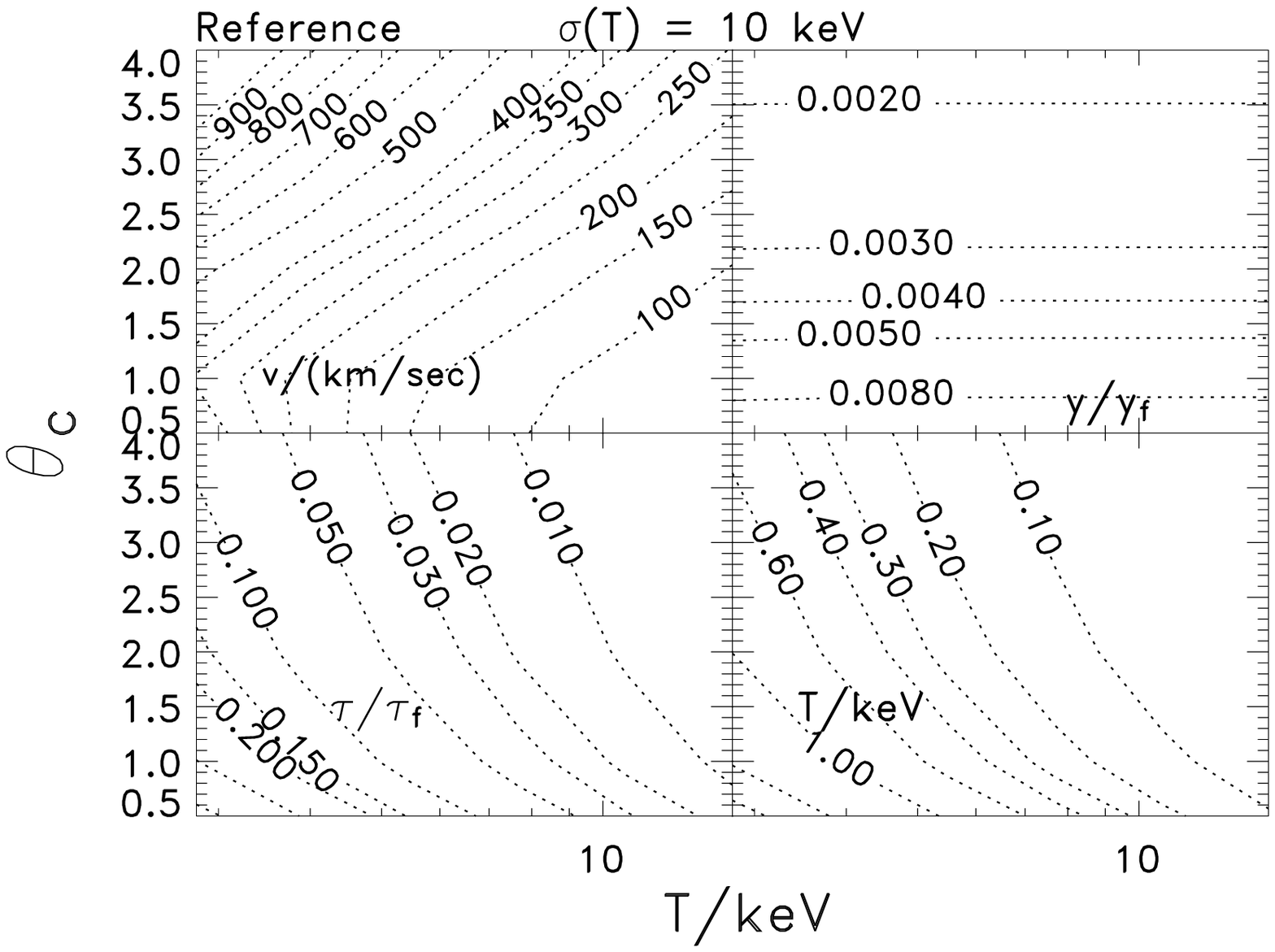}{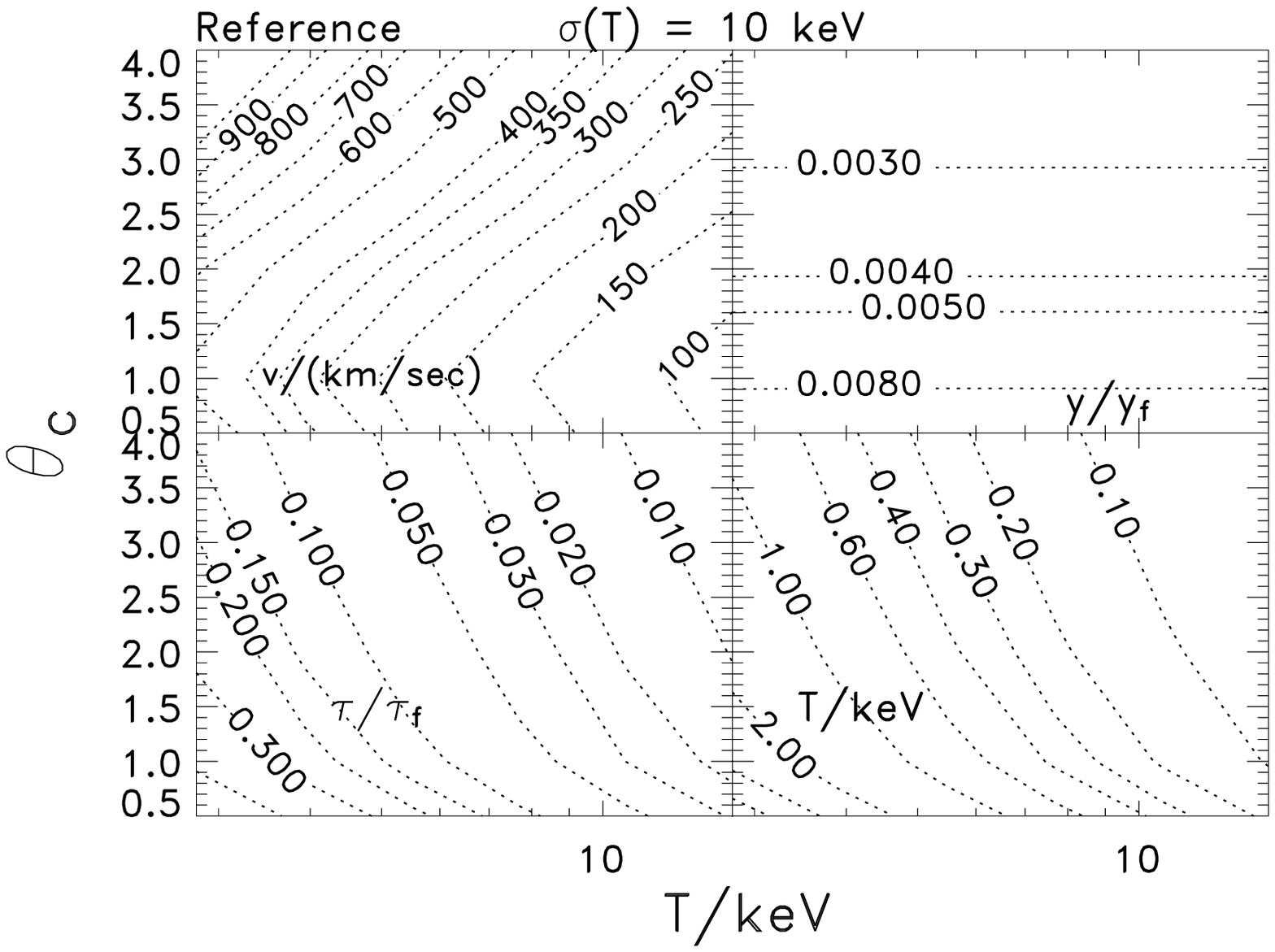}
\caption{
Forecasted uncertainties for our Reference 1', 4-band experiment
as a function of gas temperature and core radius.  On the left
we assume no point source contamination and on the right our
fiducial point source contamination model.
}
\label{fig:reference}
\end{figure*}

\subsection{Effect of Point Sources}

We begin with our fiducial cluster, studying the error forecasts
as a function of confusion noise and spectral index rms in
Fig.~\ref{fig:dusty}.  For the lower $\bar \sigma_\alpha$ values
one can infer from the flattening of the error contours with
increasing shot noise amplitude that the data themselves are
capable of fitting for the dusty galaxy component with residuals
at about the 100 to 200 $\mu$ K level.  If instrument noise were
to increase, this transition from dependence on $\langle
\left(\delta T^{\rm gal}_{350}\right)^2\rangle^{1/2}$ to
independence would occur at higher values.  The trend of error
level with $\bar \sigma_\alpha$ is as expected: flat at low
$\langle \left(\delta T^{\rm gal}_{350}\right)^2\rangle^{1/2}$ and
steeper at higher values.  Peculiar velocities are more affected,
particularly at non-zero $\bar \sigma_\alpha$, than the other
parameters.

We now fix the point source contamination parameters and study our
results as a function of galaxy cluster core radius, $\theta_c$,
and gas temperature, $T$.  Note that as we vary $\theta_c$ the map
size we assume increases in order to assure complete coverage of the
cluster and good CMB subtraction. 
For large-format bolometer arrays, expected to have more than 1000 
elements, the larger map size does not require longer integration
times because the field of view will be much larger than the galaxy
cluster. 
Therefore, despite the added coverage, one hour
of observation is still sufficient to obtain the Reference experiment
sensitivities in Table 1.
On the left of Fig.~\ref{fig:reference}
we set the point source contamination parameters to zero and on
the right to their fiducial values $\bar \sigma_\alpha = 0.3
\mu$K-arcmin and $\langle (\delta T^{\rm
gal}_{350})^2\rangle^{1/2} = 170\mu$K-arcmin.

Because of the linear response of the signal to $y$,
$\sigma(y)$ is independent of $y$ and therefore $T$ as seen in the
left side of Fig.~\ref{fig:reference}.  In contrast, at fixed $y$
and $\tau \bar \beta$ the signal is proportional to $T^2$ so one expects
$\sigma(T) \propto 1/(\partial T^2/\partial T) \propto 1/T$.  Confusion
with $y$, more important at high $T$, leads to the error decreasing more
slowly than $1/T$.  The $\sigma(\tau)$ has a similar dependence on $T$
since the negligible error on $y$ means $\sigma(\tau)/\tau = \sigma(T)/T$
and our scaling assumes $\tau \propto T$.

For fixed three-dimensional distribution of gas pressure, central
$y$ values do not vary as a function of redshift.  However, core
radii do as $\theta_c \propto 1/D_A(z)$ where $D_A(z)$ is the
angular-diameter distance to redshift $z$.
Thus to study dependence on
$z$ we vary $\theta_c$ with fixed $y$.  
Note that $D_A(z)/D_A(0.5) \simeq $
0.5, 0.7, 1.3, 1.4 for $z = 0.2$, 0.3, 1, 2 respectively, 
assuming $\Omega_m=0.3$ and $\Omega_\Lambda=0.7$.
  Note that the number of pixels with SZ signal
in any given range scales as $\theta_c^2$, so the errors on 
$\tau$, $y$ and $T$ decrease as $1/\theta_c$.

The peculiar velocities on the other hand suffer from contamination
from the CMB.  As $\theta_c$ increases it becomes more difficult to
distinguish the signal proportional to $\tau \bar \beta$ from the very
red CMB power spectrum, hence $\sigma(\bar\beta)$ goes up.  At large
$\theta_c$, where the error in $\bar \beta$ is dominated by the
CMB-induced error in $\tau \bar \beta$, the error in $\tau$ is
negligible.  In this case, $\sigma(\bar \beta) = 
(1/\tau)\sigma(\tau \bar \beta)$ so $\sigma(\beta) \propto 1/T$, 
which is the scaling we see.  At smaller $\theta_c$ uncertainty in 
$\tau$ is no longer negligible and the scaling with $T$ is more complicated.  

Including the effect of the fiducial point source contamination (right side of 
Fig.~\ref{fig:reference}) does not alter the error forecasts dramatically.
For $v$ the largest impact is at the smaller $\theta_c$ --- which is
unfortunate since this is where $v$ is measured best.  At
small $\theta_c$ the CMB noise is smaller and (less importantly) 
the lensing enhancement of confusion noise is largest.  
At $\theta_c = 30''$ the velocity errors are almost doubled.

\subsection{Effect of Varying Experimental Parameters}

\begin{figure*}
\plottwotwo{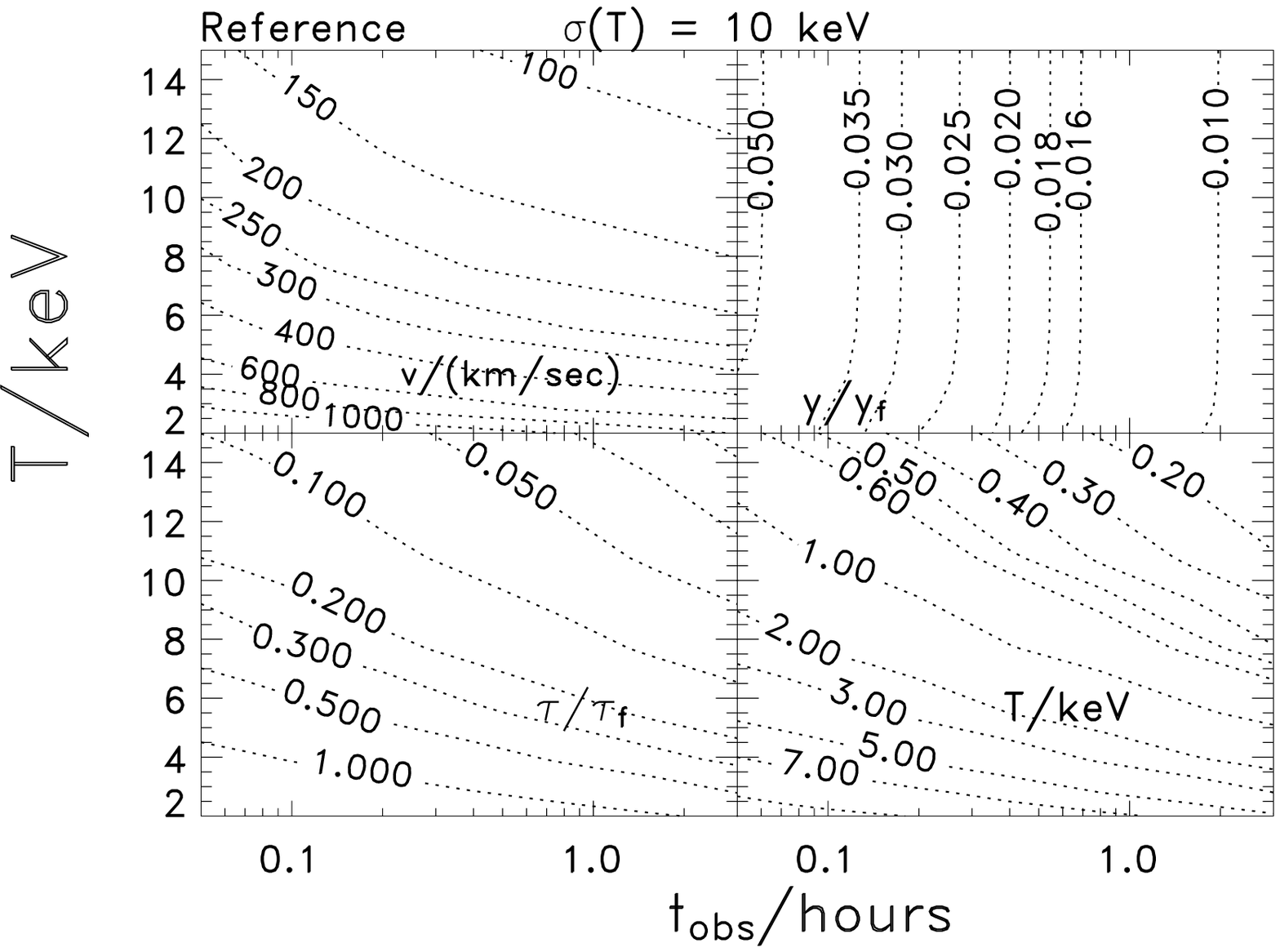}{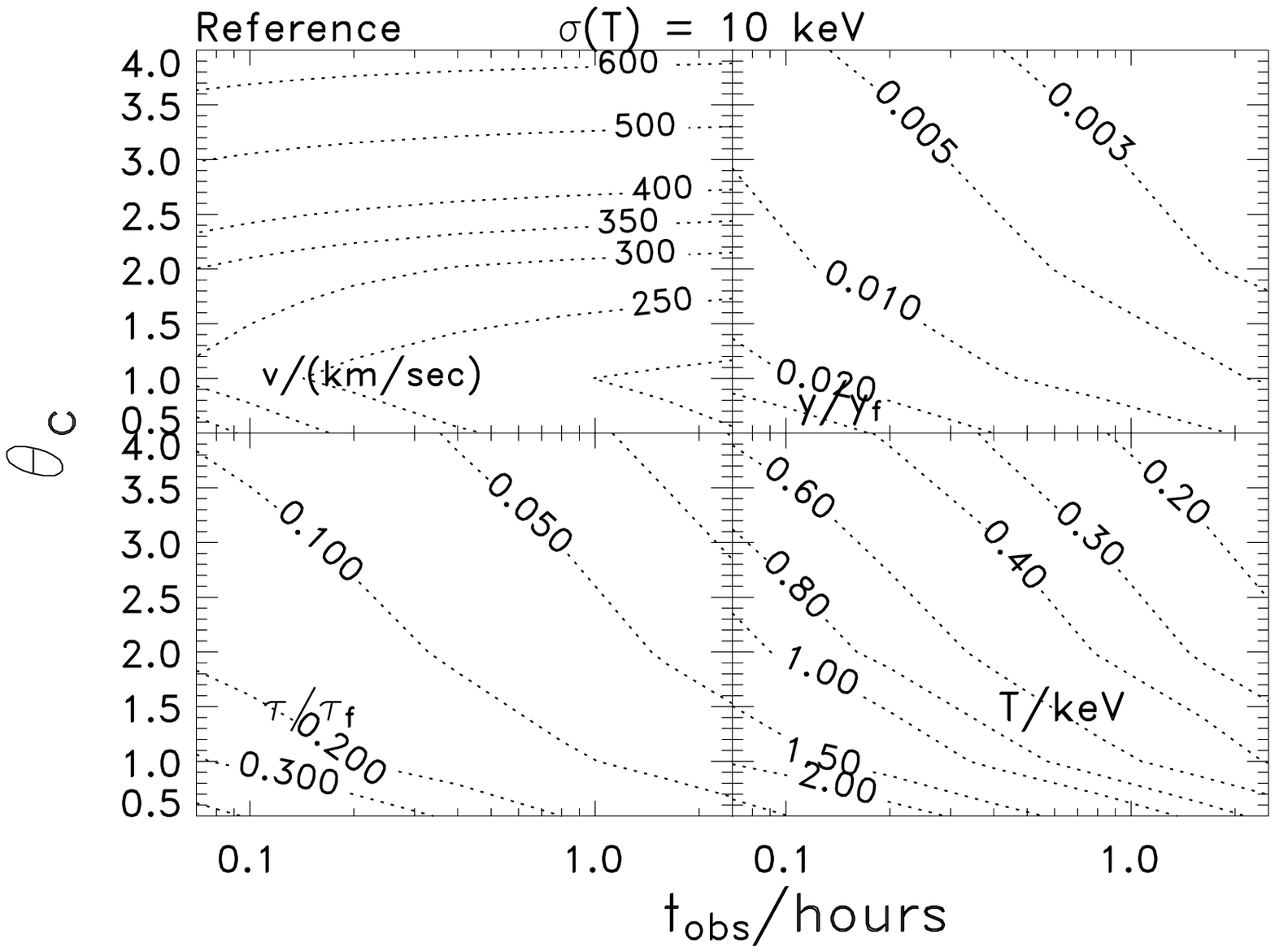}
\caption{
Forecasted uncertainties for our Reference 1', 4-band experiment
as a function of observing time and gas temperature (left) and
observing time and core radius (right).
}
\label{fig:reference_time}
\end{figure*}

Figure ~\ref{fig:reference_time} shows how the error forecasts depend
on observing time.  On the left side one can see $\sigma(y) \propto
\sqrt{1/t_{\rm obs}}$ indicating that for the Reference
experiment the dominant source of uncertainty on $y$ is instrument
noise, not dusty galaxies.  Similar scalings are seen for $\sigma(\tau)$
and $\sigma(T)$.  The $\sigma(v)$ in contrast, quickly saturate with
increased observing time producing very little improvement since
the dominant source of error is the CMB contamination.  The saturation
occurs even sooner for larger $\theta_c$.

\begin{figure}
\plotone{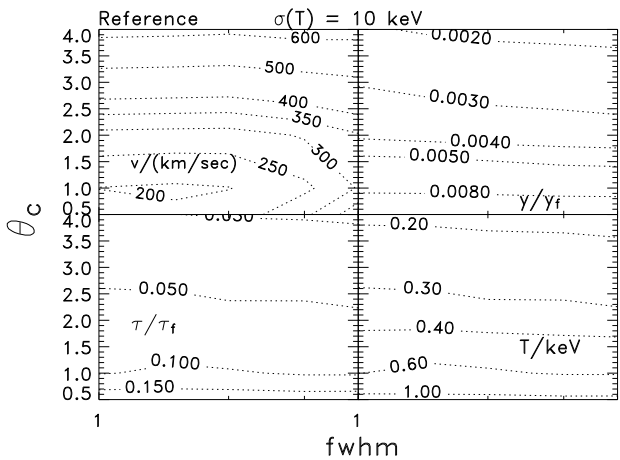}
\caption{
Forecasted uncertainties for our Reference 1', 4-band experiment
as a function of beam size and core radius.
}
\label{fig:reference_fwhm}
\end{figure}

As can be seen in Fig.~\ref{fig:reference_fwhm} parameters other than
$v$ are highly independent of the angular resolution $\theta_b$.  This
is because none of the signal is lost with increasing $\theta_b$: 
the signal is simply spread out over greater area.  
Errors would go up with increasing
$\theta_b$ for single-pixel observations as opposed to the map-making
observations we consider.  The peculiar velocity dependence on $\theta_b$
is steep for $\theta_b \ga 2 \theta_c$, since spreading out the signal
degrades the ability to subtract off the CMB contamination.  Note that to explore the effect of angular resolution independent from sensitivity, we have not scaled the sensitivity with beam size as one would expect from Eq.~\ref{model:e9}, but rather fix the weight per 
solid angle.  
The apparent decrease in $\sigma(v)$ with increasing $\theta_b$ at
fixed $\theta_c \sim 45"$ is an artifact of our sparse sampling of
the $\theta_c$, $\theta_b$ space.

\begin{figure*}
\plottwotwo{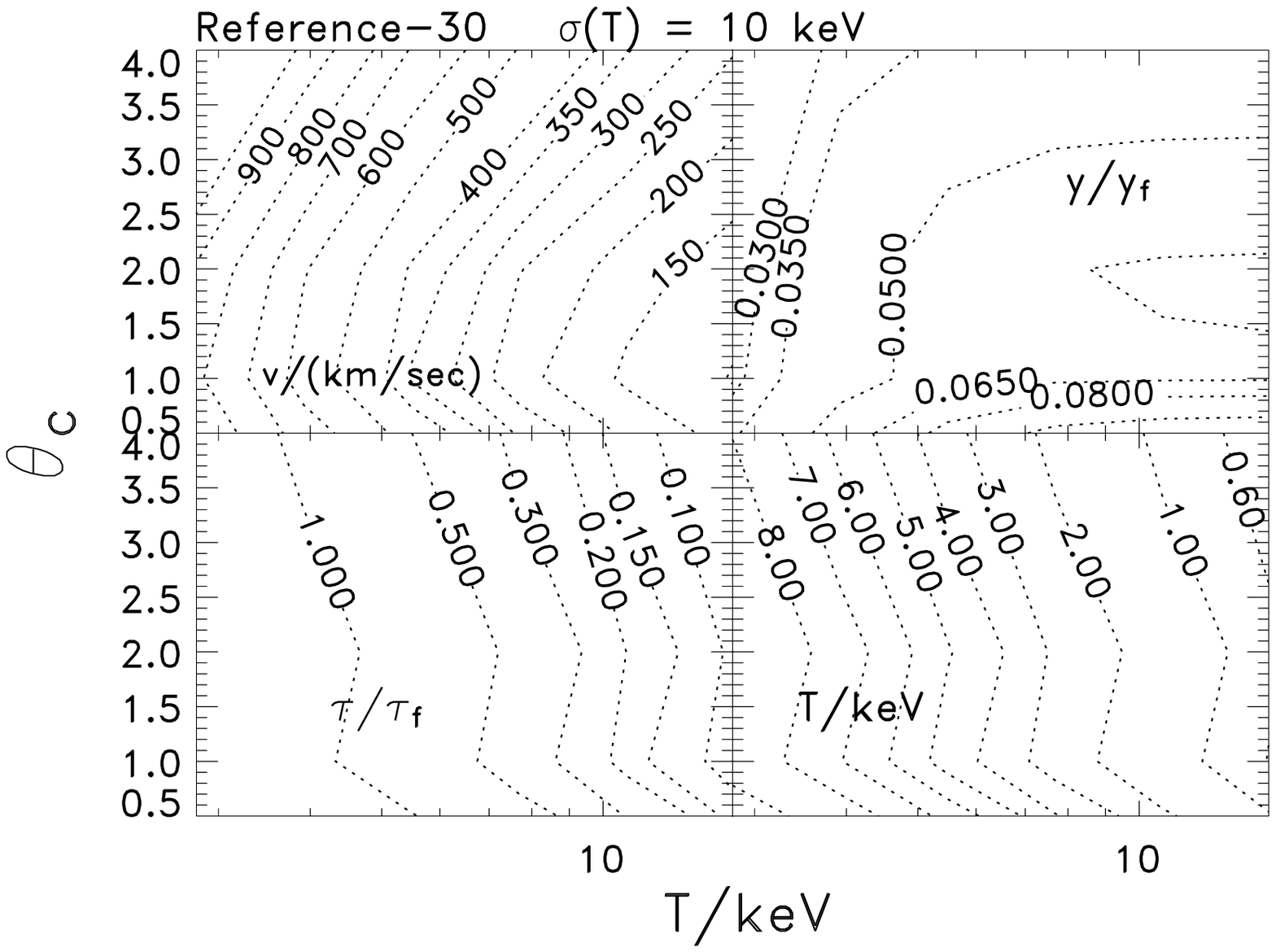}{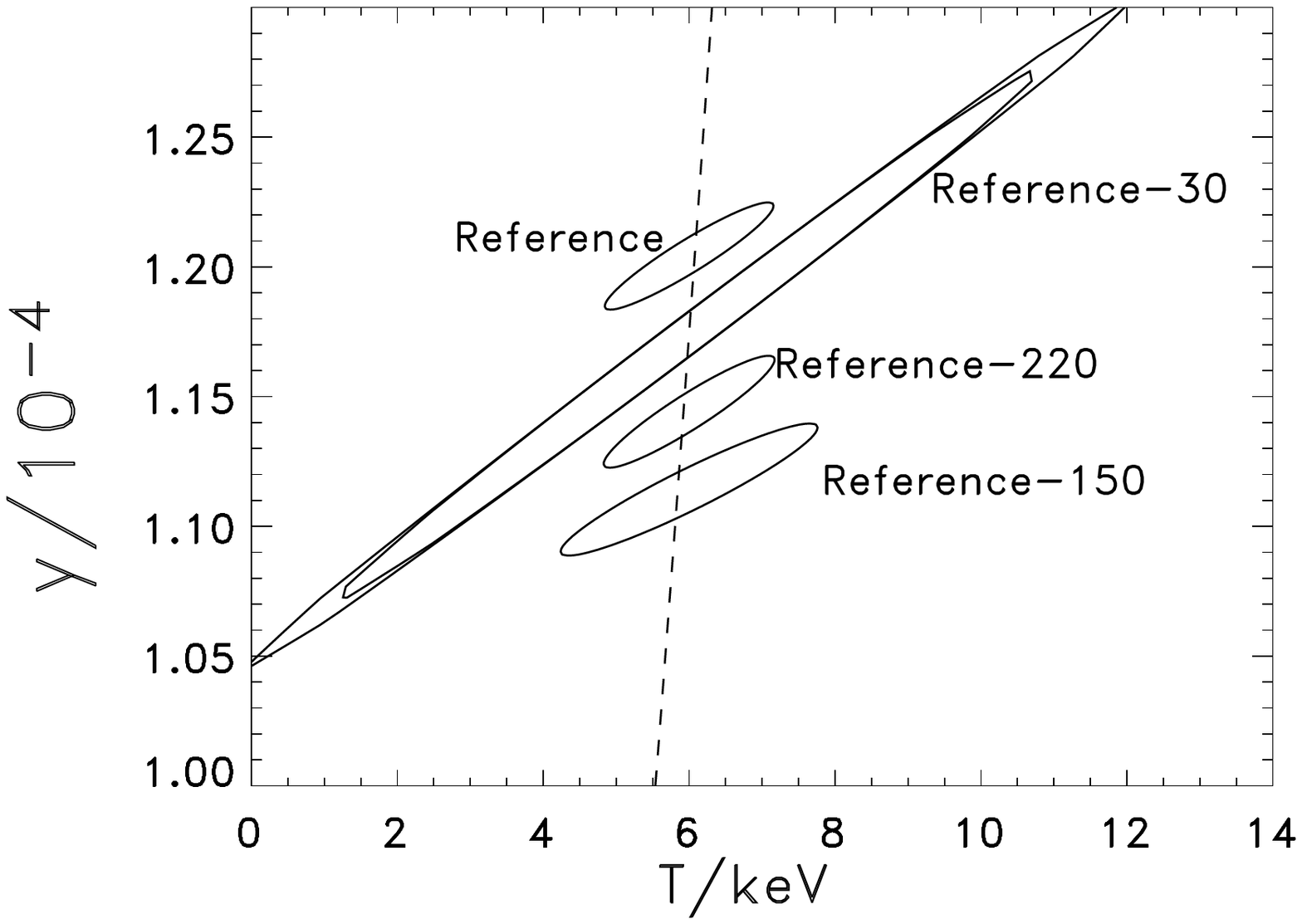}
\caption{
Forecasted uncertainties for our Reference 1', 4-band experiment
but with one channel missing.  On the left the 30 GHz channel
is the missing one.  On the right we plot contours of $y$ and $T$
for missing either no channel, 30 GHz, 150 GHz or 220 GHz.
The offsets from the fiducial value are for clarity.  The inner Reference-30 contour assumes a prior on the velocities of $\sigma_v = 300$ km/sec.
Removal of the 280 GHz channel (not
shown) makes the least difference of all. Dashed line in right panel shows
expected scaling of  McCarthy et al 2003. 
}
\label{fig:reference_minus}
\end{figure*}

We now consider the importance of each of the channels of our reference
experiment.  On the left side of Fig.~\ref{fig:reference_minus} we
show error forecasts for the Reference experiment with no 30 GHz channel, `Reference-30'.
Without this channel it is harder to distinguish a change in $y$ from a
change in $T$, thus the errors in both of these increase
considerably.  Their ratio, $\tau = y/(T/m)$, suffers similarly.  The
impact on peculiar velocities is less dramatic, especially at high
$\theta_c$ where the CMB contamination remains dominant.  Changes
are significant at smaller $\theta_c$ where, e.g., $\sigma(v)$ goes
from 250 to 400 km/sec at $T = 6$ keV.

The strong degeneracy between $y$ and $T$ for Reference-30 can be
seen on the right side of Fig.~\ref{fig:reference_minus}.  The
correlation coefficient, $r_{Ty} \equiv \langle \delta T \delta
y\rangle/(\sigma(T)\sigma(y))$ increases from 0.93 for Reference
to 0.998 for Reference-30.  This near-maximal correlation
coefficient means that the error in $T$ ($y$) is increased by a
factor of 14 compared to the error in $T$ ($y$) if $y$ ($T$) were
held fixed.  Including a reasonable prior on the velocity of 300
km/sec decreases the $T$ and $y$ errors by about 20\%. The
importance of a low frequency measurement for determining $T$ has
been emphasized by \citet{holder03} and \citet{aghanim03}.

After the 30 GHz channel, the most critical channels in
decreasing order are 150 GHz, 220 and 280.  Removal of the
150 and 220 GHz channels can increase $\sigma(T)$ by up to
a factor of 5, although only about a factor of 2 at our
small fiducial core radius of $\theta_c = 30''$ assumed in
the right side of Fig.~\ref{fig:reference_minus}.  These increases,
at the assumed level of instrument noise, are not sufficient
to degrade the peculiar velocity errors.

The dashed line on the right side of Fig.~\ref{fig:reference_minus}
is $y \propto T^2$, through our fiducial values of $y$ and $T$.
We see that the degeneracy direction is such that assuming the
scaling relation and a normalization would lead to much
more precise
determinations of both $y$ and $T$.  On the other hand, the
orientation of the degeneracy makes testing the scaling relation
more difficult than if it lay along the line.

\subsection{Planned Experiments}

\begin{figure*}
\plotthree{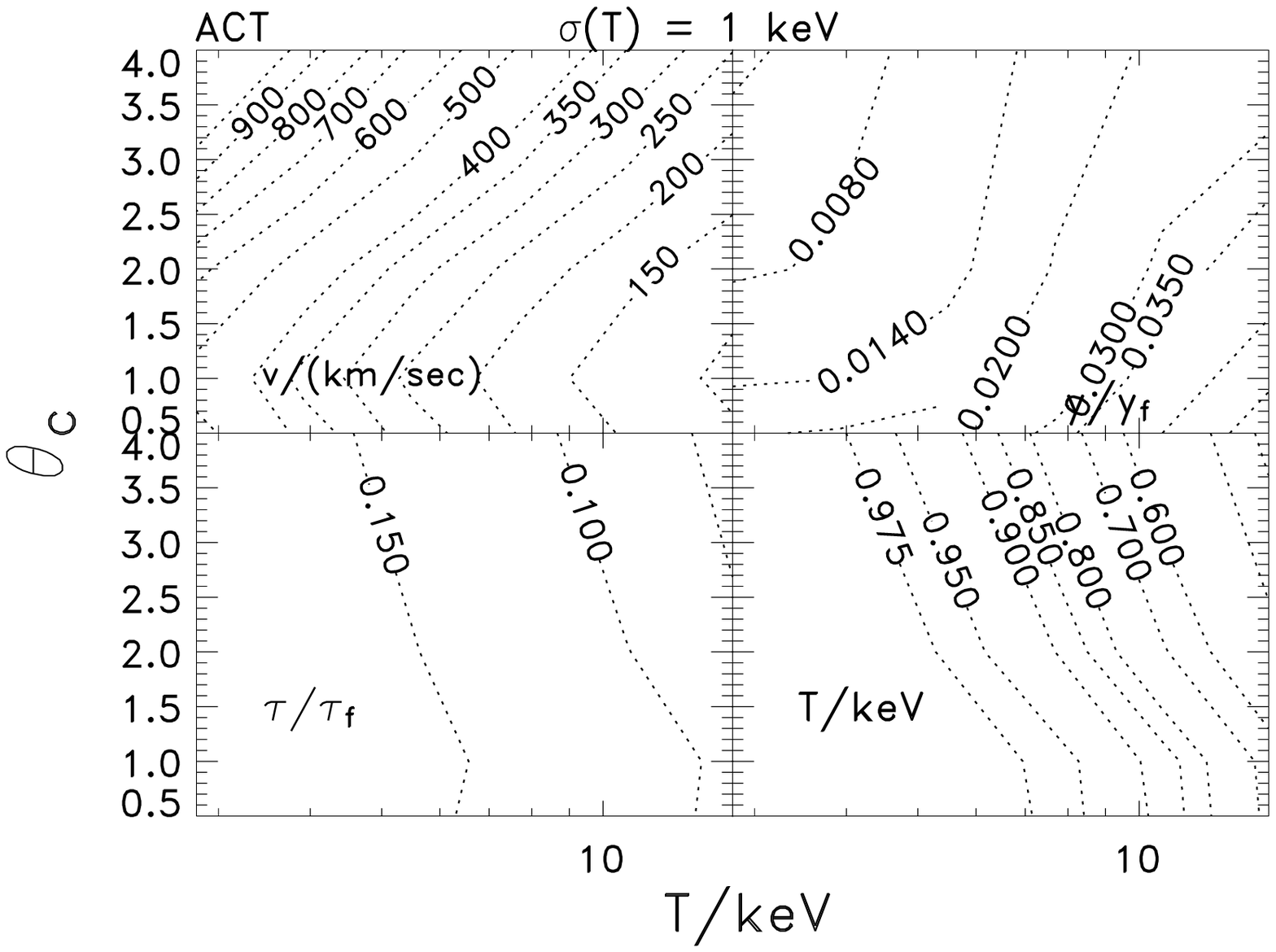}{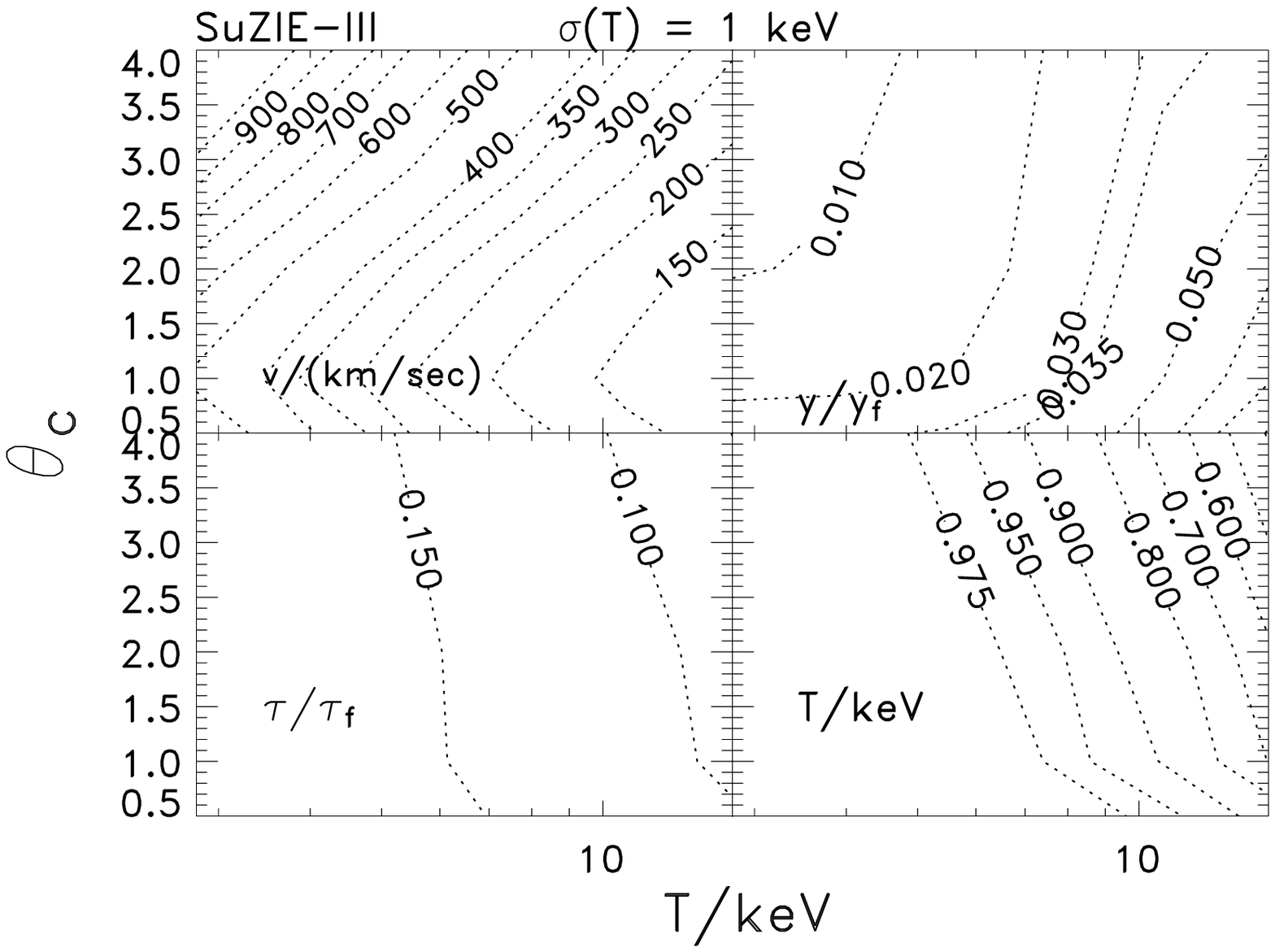}{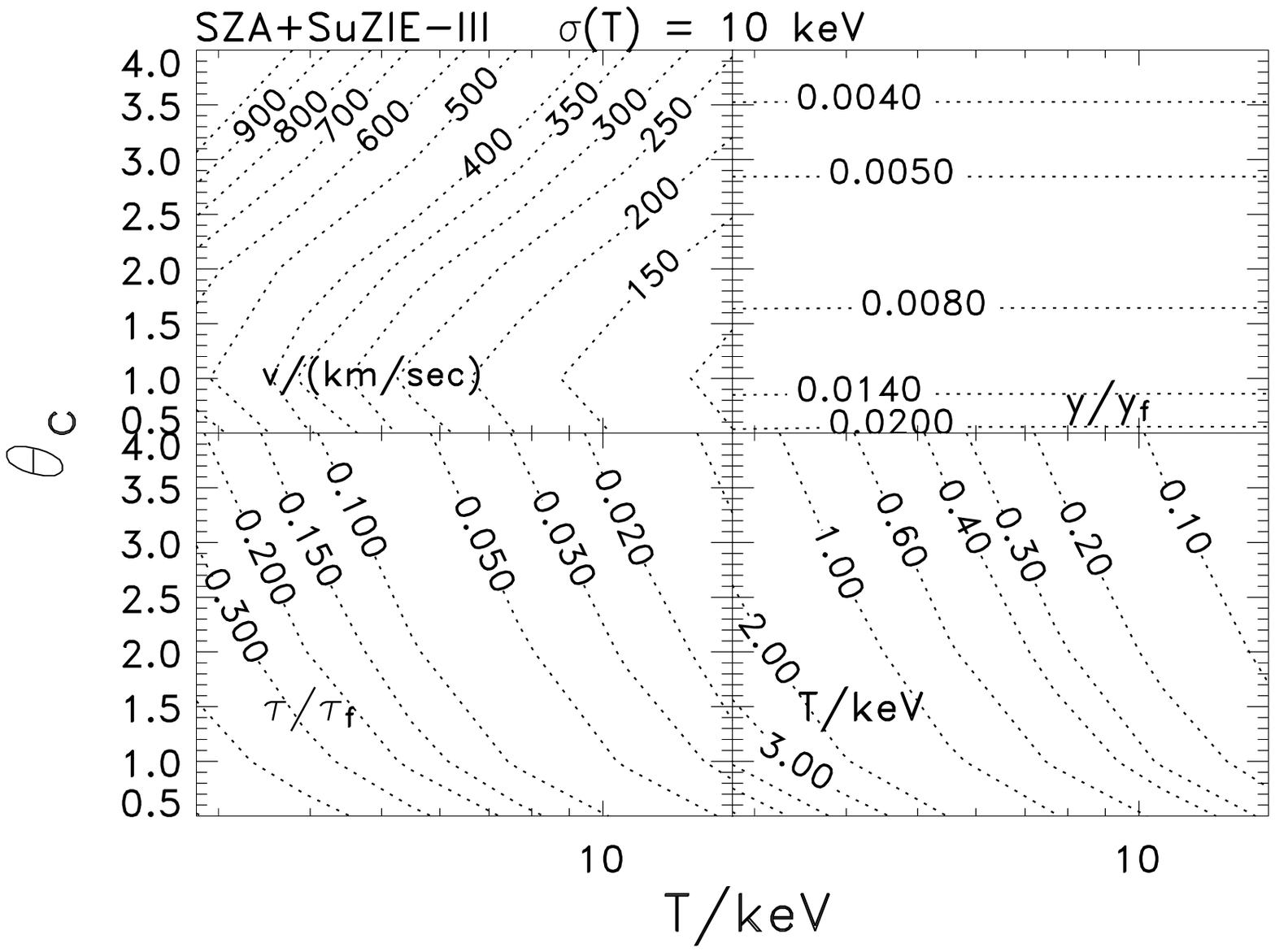}
\caption{
Forecasted uncertainties for planned experiments
as a function of gas temperature and core radius.  Left and center
are for ACT and a one-hour SuZIE-III observation respectively each
with a $\pm$1 keV determination of the temperature from X-ray data.
Right is for a combined twelve-hour SZA observation with a
one-hour per 48 sq. arcmin SuZIE-III observation.
}
\label{fig:planned}
\end{figure*}

We saw that for the Reference experiment the loss of the 30 GHz
channel increased $\sigma(T)$, as well as $\sigma(v)$ for small
$\theta_c$.  The lack of a low frequency channel can therefore be 
compensated by an X-ray determination of the temperature.  Thus we show
forecasts for the ACT survey and SuZIE-III observations both
complemented by X-ray temperature determinations of $\pm 1$ keV.  The
results for the two different experiments are very similar.  We assume one
hour of integration with SuZIE-III for every 48 sq. arcmin covered.  Recall
that in order to assure good subtraction of the CMB, we assume a map of
linear extent $\theta_M = 4(2\theta_c+\theta_b)$.  We have made no attempt
to optimize this observing strategy.  

The right-most figure in Figure~\ref{fig:planned} shows that 
adding a 30\,GHz measurement
to the millimeter data can break this degeneracy, as we expect from
our previous forecasts for the Reference experiment in 
Fig.~\ref{fig:reference}.  In 1 hour of
observation with SuZIE III (per 48 sq. arcmin of map) 
and a 12-hour SZA observation, the peculiar
velocity of a cluster can be measured to 250 km s$^{-1}$ {\em even
without an X-ray measurement of temperature}.  A low frequency
channel will thus be crucial
for obtaining peculiar velocity and gas temperatures at high
redshift where $X$-ray brightness is very low.

We also see in Figure~\ref{fig:planned} a temperature error for
our fiducial cluster of $\sim 1.5$ keV (and even smaller for
larger $\theta_c$).  
At this accuracy level, one expects to begin to see discrepancies between 
X-ray temperature measurements which are more sensitive to the hotter, 
inner regions and
SZ temperature measurements which are more sensitive to the cooler,
outer regions \citep{mathiesen01}.  Deeper integrations
can slowly improve these measurements as seen from the
study of dependence of forecasts for the Reference experiment on
observation time.

\section{Radio Point Source Contamination}

We have assumed that (i) the frequency spectrum of cluster radio
sources will be sufficiently steep at high frequencies that
measurements at 150 GHz and higher will not be contaminated and
(ii) that measurements at 30 GHz will be sufficiently high
resolution that point sources can be subtracted at that frequency
(this is possible with interferometric experiments such as the
SZA). Consequently we have concentrated only on the contamination
introduced by dusty galaxies known to be strong
submillimeter-wavelength emitters and have excluded radio sources
from our analysis. We will now justify this omission.

Clusters often contain radio point sources at mJy levels, as seen
from 30 GHz SZ measurements \citep{reese02, laroque03}. It is
important to note that radio point sources and submillimeter point
sources are {\em not} the same objects. The excess of radio
sources toward galaxy clusters is most likely due to emission from
cluster galaxies themselves \citep{cooray98}, whereas the dusty
galaxies are typically background sources that are not associated
with cluster members. Very little is known about the emission from
radio point sources at millimeter wavelengths, but their spectra
are expected to steepen due to the energy losses of the most
energetic electrons to synchotron emission. This is confirmed to
some extent by spectral indices measured between 10-90 GHz for
bright sources \citep{herbig92,sokasian01}, although there is
substantial scatter in the measured index. \citet{sokasian01} find
a mean index of $-$0.5 between 10 and 90 GHz, while
\citet{herbig92} find a mean index at 40 GHz of $-0.8\pm0.4$.
Measurements by \citet{trushkin} of the spectral indices of the
208 point sources detected by WMAP at 40 GHz
\citep{bennettfore03}, suggest a mean of $\beta = -0.1$ at 20\,GHz
with large scatter, but also show that the spectra of radio
sources are often not well-approximated by a simple power law.

Sources of only a few mJy at 30 GHz could easily still have mJy
fluxes at 150 GHz if the spectrum does not steepen significantly
above 90 GHz. 
We have used the point source counts from WMAP, DASI, VSA and CBI
summarized in \citet{bennettfore03} to determine a best fit model
of:
\begin{equation}
\frac{dN}{dS_\nu} =\frac{N_0}{S_0}
\left(\frac{S_\nu}{S_0}\right)^{-2.0}
\label{radio:e1}
\end{equation}
where $N_0 = 30$ per sq. deg. and $S_0 = 1$ mJy. 
Our expression gives number
counts for mJy sources that are slightly lower than
\citet{white03}, but that are consistent with those determined by
\citet{laroque01} based on SZ measurements of clusters at 30\,GHz.
Consequently we expect that the number of sources above 0.1 mJy at
30 GHz is less than one per 4 sq arcmin (roughly the size of a
moderate cluster). Repeating our Fisher analysis with radio
sources at this level, we find that for all the experimental
situations we have considered, the peculiar velocity uncertainty
for our canonical source is unaffected, even if a spectral index
of $\beta=0.0$ is assumed, and the number counts are 10 times
higher than we have assumed.

The spectra of mJy radio sources will soon be much better
understood when the SZA and the Array for Microwave Background
Anisotropy (AMiBA) start operation at 90 GHz. However, a complete
understanding of faint radio sources at mm wavelengths will likely
require the sensitivity and resolution of the Atacama Large
Millimeter Array (ALMA) \footnote{http://www.alma.nrao.edu}.

\section{Application of Gas Temperature and Peculiar Velocity Measurements}
\label{sec:application}

At roughly the level of 1 keV the distinction between
X-ray emission-weighted temperatures and electron-weighted
temperatures becomes important \citep{mathiesen01}, so a
comparison of the two ``gas temperatures'' could be valuable
for studies of the physics of the intra-cluster medium.

Furthermore, the pressure-weighted temperature should be an excellent
indicator of cluster mass. This could allow independent calibration
of the mass scale for galaxy cluster surveys that use X-ray, optical or
SZ selection criteria.

The measurement of peculiar velocities using the kinetic SZ effect
has been a goal of SZ measurements for many years. Peculiar velocities
at large distances will allow measurements of cosmological parameters
and reconstruction of the large scale gravitational potential.

The two-point function of radial peculiar velocities is sensitive
mainly to $\Omega_m$ and largely insensitive to the dark energy
equation-of-state parameter, $w_x$ or other cosmological
parameters. A linear theory calculation of how well the two-point
function can constrain $\Omega_m$ gives \citep{peel02} \be
\label{eqn:peelknox} \Delta \Omega_m = 0.04 \sqrt{400/N}
\frac{\langle v^2\rangle+\sigma_v^2}{\left(500 ~{\rm
km~s^{-1}}\right)^2}. \ee The SuZIE III and ACT collaborations
plan to measure optical redshifts for 400 clusters with masses
greater than $3 \times 10^{14}M_\odot$. Since typical errors for
peculiar velocities on these may be 300 km/sec and the velocity
rms is about 400 km/sec, these data may be able to achieve $\Delta
\Omega_m \simeq 0.04$.

However, Equation~\ref{eqn:peelknox} assumes that the variance in our
peculiar velocity measurement errors is perfectly well known.  In
reality, the statistical properties of the errors will be difficult to
know well because of their dependence on the subtraction of the
contamination of dusty galaxies.  If the assumed error variance is
different from the actual error variance by $f \langle v^2\rangle$
then there will be a systematic error in  $\Omega_m$ of
\be
\Delta \Omega_m = f\left(\frac{\partial \ln{\langle v^2\rangle }}{\partial\Omega_m}\right)^{-1} \simeq 0.02 (f/0.1)
\ee
independent of the number of clusters measured.  To control this
systematic error, we will need to learn more about the spectral
properties and luminosity functions of sub-mm galaxies.  To avoid this
source of systematic error one could use a linear statistic, the
average difference between the radial velocities of two clusters
separated by distance $r$ \citep{juszkiewicz99}.  How this statistic
depends on cosmological parameters is under investigation.

Another application of peculiar velocities is gravitational potential
reconstruction \citep{dekel90}.  As \citet{dore02} point out, due to
the low number density of clusters this is only possible on very large
scales.  Comparison with reconstruction from galaxy number counts can
determine galaxy biasing properties.  A good understanding of the
large scale potential would be valuable for investigating
environmental effects on galaxy formation and would be a starting
point for constrained realizations of simulations of the large--scale
structure of our universe.

\section{Discussion and Conclusions}

 Contamination by emission from
dusty galaxies will make accurate measurements of peculiar
velocities difficult. With four observing frequencies, few $\mu K$
sensitivity, and arcminute resolution it will be possible to
measure peculiar velocities at the level of roughly 200 km/s for
massive galaxy clusters.

For comparison, previous estimates of uncertainties in peculiar velocities
by Holder (2003) \nocite{holder03} assumed $\mu {\rm K}$ sensitivity
and perfect point source removal, which on Figure \ref{fig:reference}
roughly correspond to $\sigma_{\alpha}=0$ and an observing time of several
tens of hours. In that case it was found that uncertainties well below
100 km/s were possible, as could be extrapolated from Figure
\ref{fig:reference}.  Such exposure times may be prohibitively
long and
removing point source contributions to this level will likely require 
ALMA.  ALMA will provide both a better understanding of the physical nature of
dusty galaxies in general as well as measurements
of the contaminating galaxy fluxes at exactly the frequencies of interest.

The Reference experiment, as well as combinations of planned
observations, can provide measurements of peculiar velocities at
the level of 150-200 km/s, even in the presence of contamination
by dusty galaxies. This is sufficient accuracy to constrain
$\Omega_m$ to better than 10\%. At the same time galaxy cluster
temperature measurements will be possible at the keV level. This
will allow interesting comparisons with X-ray spectroscopic
measures of the X-ray emission-weighted temperature.

Peculiar velocities measured using the kinetic SZ effect, a
long-standing challenge to CMB experimentalists, may soon be a
reality. The final limiting obstacle appears to be contamination 
from dusty galaxies, but sensitive multi-frequency measurements 
should clear this hurdle.
\\

\acknowledgements{
We thank Jim Bartlett, Bradford Benson, Simona Mei and Jean-Baptiste Melin 
for useful conversations, and we thank the anonymous referee for many
constructive suggestions.  This work was supported by
the W.M. Keck foundation (GH), NASA grant NAG5-11098 (LK),
the NSF (LK) and NASA grant NAG5-12973 (SEC).
}

\end{document}